\documentclass[prd, preprint, superscriptaddress, nofootinbib, showpacs,11pt]{revtex4-1}

\usepackage{graphicx,amsmath,amsfonts,amssymb,dcolumn,epsfig,bm}
\usepackage{epsfig}%
\usepackage{graphicx}%
\usepackage{color}
\expandafter\ifx\csname package@font\endcsname\relax\else
 \expandafter\expandafter
 \expandafter\usepackage
 \expandafter\expandafter
 \expandafter{\csname package@font\endcsname}%
\fi
\newcommand\beq{\begin{equation}}
\newcommand\eeq{\end{equation}}
\newcommand\beqa{\begin{eqnarray}}
\newcommand\eeqa{\end{eqnarray}}
\newcommand\bewt{\begin{widetext}} 
\newcommand\eewt{\end{widetext}}

\begin{document}

\title{{Quantum to Classical Transition of Inflationary Perturbations}\\
 - Continuous Spontaneous Localization as a Possible Mechanism -}

\author{Suratna Das}%
\email{suratna@tifr.res.in}
\author{Kinjalk Lochan}%
\email{kinjalk@tifr.res.in}%
\author{Satyabrata Sahu}%
\email{satyabrata@tifr.res.in}%
\author{T. P. Singh}%
\email{tpsingh@tifr.res.in}
\affiliation{Tata Institute of Fundamental Research, Homi Bhabha Road, Mumbai 400005, India}

\begin{abstract}
The inflationary paradigm provides a mechanism to generate the
primordial perturbations needed to explain the observed large scale
structures in the universe. Inflation traces back all the
inhomogeneities to quantum fluctuations although the structures look
classical today. Squeezing of primordial quantum fluctuations along
with the mechanism of decoherence accounts for many aspects of this
quantum to classical transition, although it remains a matter of
debate as to whether this is sufficient to explain the issue of
realization of a single outcome (i.e. the issue of
macro-objectification) from a quantum ensemble given that the universe
is a closed system. A similar question of emergence of classical
behavior of macroscopic objects exists also for laboratory systems and
apart from decoherence there have been attempts to resolve this issue
through Continuous Spontaneous Localization (CSL), which is a
stochastic nonlinear modification of the non-relativistic
Schr\"{o}dinger equation. Recently, Martin {\it et al.} have
investigated whether a CSL-like mechanism with a constant strength
parameter, 
when the Mukhanov-Sasaki variable is taken as the 
``collapse-operator'', can explain how the primordial quantum perturbations
generated during inflation become classical. Within the scope of their
assumptions they essentially come to a negative conclusion. In the
present work, we generalize their analysis by allowing the CSL
strength parameter to depend on physical scales so as
to capture the CSL amplification mechanism. We show that such a
generalization provides a mechanism for macro-objectification
(i.e. classicalization) of the inflationary quantum perturbations,
while also preserving scale invariance of the power spectrum and phase
coherence of super-horizon perturbation modes in a particular class of
these models.  
\end{abstract} 
\maketitle
\newpage
\section{Introduction}

Inflation \cite{Guth:1980zm,Linde:1981mu} is an early phase of
accelerated expansion which apart from solving shortcomings of the Big
Bang cosmology, such as the horizon and flatness problems, also
provides a natural mechanism to seed the primordial inhomogeneities
which grow into the structures we observe today
\cite{Mukhanov:1990me,Baumann:2009ds}. The basic predictions of
inflationary paradigm such as approximate scale-invariance and
Gaussianity of primordial power spectrum are supported very well by
Cosmic Microwave Background (CMB) \cite{Hinshaw:2012fq, Ade:2013uln}
and large scale structure data \cite{Abazajian:2008wr}. Inflationary
cosmology has also been used as a testing ground of many alternative
theories like Lorentz violating theories \cite{Loretnzviolate},
modified gravity theories \cite{Modified} and more recently some
modified quantum theories \cite{Sudarsky1, Sudarsky2, Sudarsky3,
  Sudarsky4, Sudarsky5} in order to constrain them.

The mechanism of generating the primordial inhomogeneities by
inflation is essentially amplifying the quantum fluctuations of a
scalar field. But the large scale structures like galaxies do behave
like classical objects. This then leads us to the problem of
understanding quantum to classical transition in the cosmological
context which can be thought as a more serious form of the so-called
`quantum measurement problem'. The key conceptual issue of quantum
mechanics has been the appearance of deterministic outcomes in a
measurement process one performs over a quantum system prepared in a
superposed state. The quantum theory, if endowed with the Copenhagen
interpretation, suggests that a quantum state remains in a superposed
state as long as no observer does a measurement on it.  As soon as the
observer performs a measurement on the quantum state, the interaction
of the state with the classical measuring apparatus causes the quantum
state to `collapse' into one of the eigenstates of the observable
being measured by the apparatus. This age-old heuristic description has
since been drastically refined with the understanding of the
phenomenon of decoherence where the classical apparatus is coupled to
an environment consisting of a very large number of degrees of
freedom, which one is not concerned about in a measurement process
\cite{Zurek:1981xq, Zurek:1982ii, Joos:1984uk}. However, it has been
argued that the formalism of decoherence by itself does not solve the
problem of a single outcome \cite{Adler:2001us,
  Schlosshauer:2003zy}. For completion, decoherence must be
supplemented with the formalism of many-worlds \cite{Everett:1957hd},
thus making it debatable and possibly not acceptable to a section of
the community. Secondly this scheme relies heavily on a distinction
between the `environment' and the `system'. Therefore, this scheme
does not naturally extend to closed systems such as the early
universe. There have been other attempts dealing with these kind of
issues such as Bohmian mechanics \cite{Bohm:1951xw, Bohm:1951xx} but
these do not by themselves predict a testable feature of the model
which can be verified or refuted.

As has been mentioned earlier, the problem of quantum measurement
associated with the interpretation of quantum theory manifests itself
also in the context of appearance of large scale structures such as
galaxies and clusters in the universe. So the issue which becomes
pertinent is: at what stage these fluctuations lose their quantum
features to attain classical characteristics during the process of
evolution? And what is the mechanism which brings about this
quantum-to-classical transition? What we observe through CMBR is in a
sense a measurement of the field variable \cite{Kiefer:2008ku,
  Kiefer:2006je, Kiefer:1998qe} and the definiteness in the result of
the measurement indicates that the field eigenstates get selected as
the privileged basis.

Decoherence has been suggested as a plausible framework to deal with
these open problems \cite{Kiefer:2008ku, Kiefer:2006je, Kiefer:1998qe,
  Polarski:1995jg}. It turns out to be useful to work with a gauge
invariant quantity known as the Mukhanov-Sasaki variable to study the
evolution of the fluctuations.  It is interesting to note that during
the course of evolution the probability distribution of the
Mukhanov-Sasaki variable gets highly squeezed \cite{Kiefer:2008ku,
  Kiefer:2006je, Kiefer:1998qe, Polarski:1995jg, Martin:2012pe} in the
phase space along momentum. It is shown that in this high squeezing
limit the quantum expectations can be well mimicked by statistical
average over a classical stochastic field \cite{Kiefer:1998jk}. In
this sense, one can study the evolution of these fluctuations through
classical equations and take the fluctuations as classical for all
practical purposes. This is clearly one facet of explaining the
classical nature of the structures in the universe. Moreover
decoherence selects field eigenstates as the pointer basis and
eventually one arrives at a diagonal density matrix which is a
characteristic of classical systems.

However, one can still ask the question analogous to the issue of
`macro-objectification' in quantum mechanics i.e. how does a
particular realization of the system get selected.  Furthermore, from
the CMBR spectra we learn that selection of field eigenbasis as pointer
basis and the collapse of the field in one of the field eigenbasis
occurs  at least as early as the epoch of recombination since the
radiation essentially free-streams through the universe after
recombination. CMBR encodes the signature of the initial density
perturbations which started to grow after recombination. Thus the CMB
map, in a sense, carries information about a measurement of the field
variable, which took place at least as early as recombination when
there were no `conscious observers' to cause collapse.  Therefore,
 these interpretational issues, as we see, become very
  significant when applied to the early universe. Classicalization by decoherence does not naturally extend to closed
systems such as the early universe. As there were
  neither natural degrees of freedom to act as an `environment' nor
  observers who update their knowledge about the system, the schemes
  such as decoherence or state depicting the knowledge of an observer
  might seem inadequate \footnote{ However a case could be made that
  linear cosmological perturbations are not a closed system
  (i. e. there are other degrees of freedom in the primordial
  universe). Hence, decoherence can in principle be relevant for
  cosmological perturbations.} to deal with the evolution of the universe.

One possible way to approach interpretational issues such as the
quantum measurement problem is to consider dynamically induced
collapse of the wave-function. For instance, collapse may be induced
by gravity, or by modifications of quantum theory. Such modifications
will have implications for inflationary dynamics, and their role in
generating classical density perturbations has been analyzed
extensively by Sudarsky and collaborators \cite{Sudarsky1, Sudarsky2,
  Sudarsky3, Sudarsky4, Sudarsky5}.

Continuous Spontaneous Localization (CSL) is a phenomenological model
which tries to address these interpretational issues related to the
quantum theory \cite{Bassi:2003gd, Bassi:2012bg}. In this scheme one
makes non-linear, non-hermitean stochastic modifications to the
evolution equations otherwise governed by the hermitean
Hamiltonian. The defining feature of these corrections is that they
drive the state to one of the eigenstates of position (in the position
driven models) which is a description of localization. The quantum
measurement process is understood in this framework with the
understanding that any measurement is essentially a position
measurement \cite{Bassi:2003gd, Bassi:2012bg}. One important aspect of
the modifications is that they are very tiny for microscopic systems
resulting in the longevity of the superposed states, while they grow
substantially large for macroscopic systems (the so-called
amplification mechanism) causing a quick and effective position
localization of macroscopic objects. Being a phenomenological
modification of quantum theory, its parameters are constrained through
studying the effects of such models on laboratory as well as
cosmological situations \cite{Bassi:2003vf, Adler:2004un,
  Adler:2004rf,Pearle:2007rw, Pearle:2010uu, Lochan:2012di}. For a
detailed introduction of this scheme the reader is referred to
\cite{Bassi:2003gd, Bassi:2012bg}.

Although this scheme is phenomenological and only developed for
non-relativistic systems, motivated by the features of this scheme
Martin {\it et al.}  \cite{Martin:2012pe} recently attempted a
generalization of adding stochastic modification to quantum theory to
study the evolution of the very early universe. As a relativistic
generalization of CSL has not yet been successfully achieved
\cite{Pearle:2005rc, Pearle:1976ka, Ghirardi:1985mt, Pearle:1988uh,
  Ghirardi:1989cn, Bassi:2003gd, Weinberg:2011jg}, in this case they
introduce a CSL-like correction in the functional Schr\"{o}dinger
equation for evolution of mode functions. They show that this analysis
is precisely equivalent to consideration of a harmonic oscillator with
time dependent frequency in the original CSL model.  The motivation is
to obtain classicalization for super-horizon modes in a single
realization of the universe.
However, for simplicity they primarily consider a
model with a constant collapse strength parameter
$\gamma$ in Mukhanov-Sasaki variable driven collapse mechanism
which lacks the mechanism of scale-dependent amplification as we will discuss below.
The amplification mechanism in the non-relativistic collapse model (which assumes $\gamma$ to be
mass-dependent) causes the non-linear stochastic corrections to become
dominant for macro-objects causing an effective and efficient
collapse. The outcome of their analysis with constant $\gamma$ model
is that one does not obtain an efficient collapse of the field
variable. Moreover, they show that such modifications leave
undesirable imprints on the power spectrum by making it
scale-dependent. The requirement that these modes must reside outside
the current cosmological horizon, motivated by the fact that the power
spectrum we observe is scale invariant, constrains the parameter
$\gamma$ to exponentially suppressed values. Although the time
available for collapse process is sufficient with the constrained
parameter, the suppressed value of the CSL correction parameter
$\gamma$ makes the collapse inefficient as the width of localization
is inversely related to $\gamma$.

This scheme adopted by Martin {\it et al.} although certainly being a
preliminary and interesting step towards obtaining the guiding
principles for a field theoretic generalization of CSL, has some
limitations, in our opinion. The first and obvious one, as pointed out
by them, is the lack of an amplification mechanism dependent on the 
relevant length scales. As discussed
before, the amplification mechanism is the source of different
behavior of macro or micro-objects. One in general, does not expect an
efficient collapse in the absence of this mechanism.  Had the collapse
process without the amplification been still effective, one would be
lead to the conclusion that modes with smallest length scales are as
effectively classicalized as modes with large length scales rendering
any quantum phenomenon for any length scale. Therefore, it will be a
natural demand to ask for a CSL generalization which calls for a
distinction between micro and macro modes\footnote{ 
 Also, Martin {\it et al.} argue that there exists an ambiguity in selecting
the collapse-operator at the phenomenological level. In their analysis they 
primarily take the Mukhanov-Sasaki (MS) variable as the collapse-operator with 
constant strength parameter. Using the ambiguity in defining the collapse-operator
they show that a time dependent collapse
operator with constant strength parameter is equivalent to a time-dependent strength 
parameter driven collapse by the MS variable.
It can be argued that the collapse can be $\hat{\zeta}_k$  (MS variable) driven or
$h(a)\hat{\zeta}_k$ driven (for some arbitrary function $h(a)$ of the scale factor $a$). Both of these operators
drive the field to an eigenstate of $\hat{\zeta}_k$, with same probabilities.
 If we take the collapse being driven by  $h(a)\hat{\zeta}_k$ with a constant $\gamma$, the analysis
will be equivalent to a collapse process driven by $\hat{\zeta}_k$ with a time-dependent $h^2(a)\gamma$.
Thus the ambiguity in selecting the collapse operator translates into a time-dependent 
$\gamma$ analysis.
}.

Secondly, one learns that such a scheme is capable of distorting the
scale invariant power spectrum for the modes where the CSL correction
is dominating over the standard quantum evolution. Furthermore, in
normal CSL scenario one expects an efficient collapse to occur in the
above mentioned regime. Thus when one suggests that the modes which
distort the power spectrum lie outside the current horizon, one is
lead to the argument that the modes physically relevant to us and
within the horizon today are the ones least affected by CSL
modifications. However these are the modes which one sought to be
classicalized by the CSL mechanism, i.e. dominated by the CSL
modifications. Thus one should ask for a mechanism which not only
captures the amplification mechanism but also respects the observed
scale invariant power spectrum.

Our scheme in this paper will be to constrain the class of CSL type
modifications in the spirit of Martin {\it et al.} by consideration of
the above mentioned criteria. We will see that the inefficiency of the
constant $\gamma$ models is manifested at the level of Wigner function
itself, even before constraining $\gamma$ to a insignificantly small
value when confronted with observations. As argued before, to take
into account this discrepancy we will modify the CSL parameter
$\gamma$ to carry information about different modes. We show that any
time invariant model is as inefficient as a constant $\gamma$
case for MS variable driven collapse. Therefore, we introduce scale
dependence in this parameter. In
such a case, we observe that unless the scale dependence of $\gamma$ is
sufficiently strong there is no localization in the field variable. If
the parameter lies in a certain allowed range then we obtain desired
effective collapse in field eigenbasis in the superhorizon limit. We
further constrain the model by considering the scale-invariance of the
power spectrum. We show that a particular class of scale dependent
models seems capable of meeting these requirements. Further, the
appearance of acoustic peaks in the CMBR map is a signature of phase
coherence of primordial perturbations which appears very naturally in
standard inflationary scenarios \cite{Dodelson:2003ip,
  Albrecht:1995bg}. In the class of collapse models which respect the
scale-invariance of power spectrum we argue that the phase coherence
is not destroyed.

The paper is organized as follows. We briefly introduce the
inflationary perturbations in the standard framework in the Heisenberg
as well as Schr\"{o}dinger picture in section II. Here we also discuss
the concept of squeezing and the understanding of classicalization
through it. In section III, we briefly review the CSL model and
discuss the models of constant $\gamma$ (as well as $\gamma(k)$) in
inflationary context to study its effects on squeezing and power
spectrum. Section IV describes the introduction of amplification
mechanism in the form of scale dependent $\gamma$ model. We discuss its
effect on power spectrum and constrain the models which respect the
scale invariance of the power spectrum. In section V, we discuss the
issue of phase coherence in the light of the CSL type modification. We
then summarize our main results in section VI and conclude.


\section{Inflationary Perturbations in Heisenberg and Schr\"{o}dinger picture and squeezed states}

In general, the perturbations generated during inflation are studied
in the Heisenberg picture \cite{Mukhanov:1990me}. We, on the other
hand, need to study the evolution of these primordial perturbations in
the Schr\"{o}dinger picture: this will help us incorporate the CSL
mechanism within the arena of inflation. These two representations of
quantum fluctuations are equivalent and provide the same physical
implications for the derived quantities.  As the Schr\"{o}dinger
picture of evolution of primordial fluctuations is less studied in the
literature, we will recall here the evolution of mode functions in a
generic inflationary scenario (especially the squeezing of the modes)
in both the pictures; this will help us relate the more conventional
Heisenberg picture analysis of the inflationary modes with that of the
Schr\"{o}dinger picture.

\subsection{Heisenberg Picture}

We first provide a brief account of the inflationary perturbations and
the squeezing of the modes in the Heisenberg picture, following
\cite{Polarski:1995jg}. Considering only the scalar fluctuations of a
perturbed FRW metric in conformal time $\tau$
\begin{eqnarray}
  ds^2=a^2(\tau)\left[-(1-2A)d\tau^2+2(\partial_iB)dx^id\tau+\left\{(1-2\psi)\delta_{ij}+2\partial_i\partial_jE\right\}dx^idx^j\right],
\end{eqnarray}
one can combine the fluctuations of the inflaton field
$\varphi(\tau,\mathbf{x})\equiv\varphi_0(\tau)+\delta\varphi(\tau,\mathbf{x})$
and the scalar degrees of freedom of the perturbed FRW metric to
construct gauge-invariant quantities as
\begin{eqnarray}
\Phi_B(\tau,\mathbf{x})&=&A+\frac1a\left[a(B-E')\right]',\\
\delta\varphi^{\rm gi}(\tau,\mathbf{x})&=&\delta\varphi+\varphi_0'(B-E'),
\end{eqnarray}
where $x'\equiv {dx}/{d\tau}$ and $\Phi_B$ is known as the
Bardeen potential. These two gauge-invariant quantities, $\Phi_B$
and $\delta\varphi^{\rm gi}$, are related to each other by perturbed
Einstein equations. A combination of these two gauge-invariant
quantities, known as the Mukhanov-Sasaki (MS) variable
\begin{eqnarray}
\zeta(\tau,\mathbf{x})=a\left[\delta\varphi^{\rm gi}+\varphi_0'\frac{\Phi_B}{\mathcal{H}}\right],
\end{eqnarray}
where $\mathcal{H}={a'}/{a}$, is often studied in the context of
evolution of primordial fluctuations, because in the absence of anisotropic
stress in the energy-momentum tensor the MS variable is related to the
comoving curvature perturbation $\mathcal{R}$:
\begin{eqnarray}
\zeta(\tau,\mathbf{x})=\frac{a\varphi_0'}{\mathcal{H}}\mathcal{R}(\tau,\mathbf{x}).
\label{ms-R}
\end{eqnarray}
The curvature perturbation remains conserved on super-horizon scales. 

By treating it as a field, the action of
$\zeta(\tau,\mathbf{x})$ (expanding up to second order of
perturbations) can be written as \cite{Mukhanov:1990me}
\begin{eqnarray}
S=\frac12\int d^4x \left[(\zeta')^2-\left(\nabla\zeta\right)^2+\frac{z''}{z}\zeta^2\right],
\label{action1}
\end{eqnarray}
where $z\equiv {a\varphi_0'}/{\mathcal{H}}=a\sqrt{2\epsilon}M_{\rm
  Pl}$, $\epsilon$ being the slow-roll parameter and $M_{\rm Pl}$
being the reduced Planck mass. This action is equivalent to the action
\begin{eqnarray}
S=\frac12\int d^4x \left[(\zeta')^2-\left(\nabla\zeta\right)^2-2\frac{z'}{z}\zeta\zeta'+\left(\frac{z'}{z}\right)^2\zeta^2\right]
\label{action2}
\end{eqnarray}
up to a total derivative term \cite{Albrecht:1992kf}. 

In this paper we  consider the slow-roll parameter $\epsilon$ to
vary negligibly with time during inflation; this will correspond to
${z''}/{z}={a''}/{a}$ and ${z'}/{z}={a'}/{a}$. We
 also assume a quasi-de Sitter spacetime for studying the evolution
of modes during inflation in which case one can write the scale factor
as
\begin{eqnarray}
a(\tau)=-\frac{1}{H\tau(1-\epsilon)},
\end{eqnarray}
where $H$ is the Hubble parameter. One then gets
${a'}/{a}\approx-(1+\epsilon)/\tau$ up to first order in
slow-roll parameter $\epsilon$. For convenience we  neglect the
slow-roll parameter (as $\epsilon\ll1$ for quasi-de Sitter space) to
keep the leading order term and consider
${a'}/{a}=-1/\tau$ for all practical purposes.

Initially we will consider this action for analysis of squeezing of
modes. Treating the scalar fluctuations classically, the conjugate
momentum $p$ of the MS variable $\zeta(\tau,\mathbf{x})$ would be
\begin{eqnarray}
p\equiv \frac{\partial \mathcal{L}(\zeta,\zeta')}{\partial \zeta'}=\zeta'-\frac{a'}{a}\zeta.
\end{eqnarray}
Decomposing the MS variable in Fourier modes:
\begin{eqnarray}
\zeta(\tau,\mathbf{x})=\frac{1}{(2\pi)^{\frac32}}\int d^3{\mathbf k}\,\zeta_{\mathbf k}(\tau)e^{i\mathbf{k\cdot x}},
\end{eqnarray}
with $\zeta_{-\mathbf k}=\zeta^*_{\mathbf k}$ as $\zeta(\tau,\mathbf{x})$
is real, the Hamiltonian of the system is given by 
\begin{eqnarray}
H\equiv\int d^3{\mathbf x}\,\mathcal{H}=\frac12\int d^3{\mathbf k}\left[p_{\mathbf k}p^*_{\mathbf k}+k^2\zeta_{\mathbf k}\zeta^*_{\mathbf k}+\frac{a'}{a}\left(\zeta_{\mathbf k}p^*_{\mathbf k}+\zeta^*_{\mathbf k}p_{\mathbf k}\right)\right]
\label{class-H}
\end{eqnarray}
and the modes $\zeta_{\mathbf k}$ satisfy the equation of motion 
\begin{eqnarray}
\zeta_{\mathbf k}^{\prime\prime}+\omega^2(\tau,k)\zeta_{\mathbf k}=0,
\label{eom}
\end{eqnarray}
which is equivalent to an equation of motion of a harmonic oscillator
with a time-dependent frequency
\begin{eqnarray}
\omega^2(\tau,k)=k^2-\frac{a''}{a}.
\label{omega}
\end{eqnarray}

Upon quantization of the classical field $\zeta(\tau,\mathbf{x})$, the
Fourier transforms $\zeta_{\mathbf k}$ are promoted to operators as
\begin{eqnarray}
\hat\zeta_{\mathbf k}=\frac{a_{\mathbf k}+a^\dagger_{-\mathbf k}}{\sqrt{2k}}\,,\quad\quad \hat p_{\mathbf k}=-i\sqrt{\frac k2}\left(a_{\mathbf k}-a^\dagger_{-\mathbf k}\right),
\end{eqnarray}
and the canonical commutation relations
\begin{eqnarray}
\left[\hat\zeta(\tau,\mathbf{x}), \hat p(\tau,\mathbf{y})\right]=i\delta^3(\mathbf{x}-\mathbf{y})
\label{commute}
\end{eqnarray}
provide the commutation relations in the Fourier space as 
\begin{eqnarray}
\left[\hat \zeta(\tau,\mathbf{k}), \hat p^\dagger(\tau,\mathbf{k'})\right]&=&i\delta^3(\mathbf{k}-\mathbf{k'}),\nonumber\\
\left[a(\tau,\mathbf{k}), a^\dagger(\tau,\mathbf{k'})\right]&=&\delta^3(\mathbf{k}-\mathbf{k'}).
\label{commute-fourier}
\end{eqnarray}
Hence the classical Hamiltonian given in Eqn.~(\ref{class-H}) yields
the Hamiltonian operator:
\begin{eqnarray}
\hat H=\frac12\int d^3{\mathbf k}\left[k\left\{a_{\mathbf k}a^\dagger_{\mathbf k}+a^\dagger_{-\mathbf k}a_{-\mathbf k}\right\}-i\frac{a'}{a}\left\{a_{\mathbf k}a_{-\mathbf k}-a^\dagger_{\mathbf k}a^\dagger_{-\mathbf k}\right\}\right].
\end{eqnarray}
For the above Hamiltonian one gets the time evolution of the creation
and annihilation operators as
\begin{eqnarray}
\left(\begin{array}{c}
a'_{\mathbf k}\\
a^{\dagger\prime}_{-\mathbf k}
\end{array}\right)=\left(\begin{array}{ccc}
-ik&\quad&\frac{a'}{a}\\
\frac{a'}{a}&\quad&ik
\end{array}\right)\left(\begin{array}{c}
a_{\mathbf k}\\
a^{\dagger}_{-\mathbf k}
\end{array}\right).
\end{eqnarray}
A general solution for the above coupled equations is 
\begin{eqnarray}
a_{\mathbf k}(\tau)&=&u_k(\tau)a_{\mathbf k}(\tau_0)+v_k(\tau)a^\dagger_{-\mathbf k}(\tau_0),\nonumber\\
a^\dagger_{-\mathbf k}(\tau)&=&u^*_k(\tau)a^\dagger_{-\mathbf k}(\tau_0)+v^*_k(\tau)a_{\mathbf k}(\tau_0),
\end{eqnarray} 
which yields the commutation relation given in
Eqn.~(\ref{commute-fourier}) provided
\begin{eqnarray}
|u_k(\tau)|^2-|v_k(\tau)|^2=1.
\label{constraint}
\end{eqnarray}
This also shows that $u_k$ and $v_k$ follow the evolution equations
\begin{eqnarray}
u_k'&=&-iku_k+\frac{a'}{a}v_k^*\,,\nonumber\\
v_k'&=&-ikv_k+\frac{a'}{a}u_k^*\,.
\end{eqnarray}

One can also write the Fourier transform $\zeta_{\mathbf k}$ in terms
of mode functions $f_k(\tau)$ as
\begin{eqnarray}
\hat\zeta_{\mathbf k}=f_k(\tau)a_{\mathbf k}(\tau_0)+f^*_k(\tau)a^\dagger_{-\mathbf k}(\tau_0),
\label{zetak}
\end{eqnarray}
which satisfies the commutation relation given in Eqn.~(\ref{commute})
provided that the conserved Wronskian satisfies
\begin{eqnarray}
W=f_kf_k^{*\prime}-f_k^*f_k'=i.
\end{eqnarray}
The mode function also satisfies the Euler-Lagrange equation given
in Eqn.~(\ref{eom}):
\begin{eqnarray}
f_k''+\omega^2(\tau,k)f_k=0
\label{mode-eom}
\end{eqnarray}
and is related to $u_k$ and $v_k$ as
\begin{eqnarray}
f_k=\frac{u_k+v_k^*}{\sqrt{2k}}.
\label{f-uv}
\end{eqnarray}
We also note here that $f_k^*$, too, satisfies the second-order
differential equation stated above and thus $f_k$ and $f_k^*$ are the
two linearly independent solutions of the evolution equation of the
mode functions. This is also evident from the Wronskian of $f_k$ and
$f_k^*$ being non-zero. For the momentum modes we have
\begin{eqnarray}
\hat p_{\mathbf k}=-i\left[g_k(\tau)a_{\mathbf k}(\tau_0)-g^*_k(\tau)a^\dagger_{-\mathbf k}(\tau_0)\right],
\label{pk}
\end{eqnarray}
which is related to $u_k$ and $v_k$ as
\begin{eqnarray}
g_k=\sqrt{\frac k2}(u_k-v_k^*).
\label{g-uv}
\end{eqnarray}

We can study the evolution of the modes in terms of another set of
variables $r_k$, $\theta_k$ and $\phi_k$ which are used in the
so-called squeezed state formalism \cite{Albrecht:1992kf}. Using the
constraint given in Eqn.~(\ref{constraint}) the old variables
$u_k$ and $v_k$ can be reparametrized as
\begin{eqnarray}
u_k(\tau)&=&e^{-i\theta_k(\tau)}\cosh r_k(\tau),\nonumber\\
v_k(\tau)&=&e^{i\theta_k(\tau)+2i\phi_k(\tau)}\sinh r_k(\tau),
\end{eqnarray}
where $r_k$ and $\phi_k$ are the squeezing parameter and squeezing
angle respectively and $\theta_k$ is the phase. The evolution equation
of these three parameters would be
\begin{eqnarray}
r_k'&=&\frac{a'}{a}\cos2\phi_k,\nonumber\\
\phi_k'&=&-k-\frac{a'}{a}\sin 2\phi_k\coth 2r_k,\nonumber\\
\theta_k'&=&k+\frac{a'}{a}\sin 2\phi_k\tanh r_k.
\label{rtp-eom}
\end{eqnarray}
The above equations suggest that when $|r_k|\rightarrow\infty$ (which
happens on super-horizon scales as we see later in this section) one
gets $(\theta_k+\phi_k)'|_{|r_k|\rightarrow\infty}=0$. This shows that
on super-horizon scales $\theta_k+\phi_k\rightarrow\delta_k$ where
$\delta_k$ is some constant phase. As on super-horizon scales the mode
function becomes (following Eqn.~(\ref{f-uv}))
\begin{eqnarray}
f_k|_{|r_k|\rightarrow\infty}\rightarrow e^{-i\delta_k}e^{r_k}\cos\phi_k,
\end{eqnarray}
it shows that the phase of the modes becomes constant on super-horizon
scales. Another way of showing that the phase of the mode functions
freezes on super-horizon scales is by writing the mode functions as 
\begin{eqnarray}
f_k(\tau)=R_k(\tau)e^{-i\delta_k(\tau)},
\label{fRdelta}
\end{eqnarray}
which yields the evolution equation of the amplitude $R_k$ and phase
$\delta_k$ as
\begin{eqnarray}
R_k''+(\omega^2-\delta_k^{'2})R_k&=&0,\nonumber\\
\delta_k''+2\frac{R_k'}{R_k}\delta_k'&=&0.
\label{phase-coherence}
\end{eqnarray}
The second equation shows that $\delta_k'=0$ is a fixed point solution
of this equation and this also shows that the amplitude $R_k$ follows
the same evolution equation as $f_k$ on super-horizon scales. We will
see in more detail in Sec.~\ref{ph-coh}, how this freezing of phases,
in standard inflation, is related to the appearance of acoustic peaks
in the CMBR.

Also, the solutions for the evolution equation of mode functions
for a massless scalar are
\begin{eqnarray}
f_k&=&\frac{1}{\sqrt{2k}}\left(1-\frac{i}{k\tau}\right)e^{-ik\tau},\nonumber\\
g_k&\equiv&i\left(f_k'-\frac{a'}{a}f_k\right)=\sqrt{\frac k2}e^{-ik\tau}, \label{f-ksol}
\end{eqnarray}
which, after some straightforward calculations, yield
\begin{eqnarray}
r_k&=&\sinh^{-1}\left(\frac{1}{2k\tau}\right),\nonumber\\
\phi_k&=&\frac\pi4-\frac12\tan^{-1}\left(\frac{1}{2k\tau}\right),\nonumber\\
\theta_k&=&k\tau+\tan^{-1}\left(\frac{1}{2k\tau}\right)
\label{rtp-sol}
\end{eqnarray}
and satisfy the set of evolution equations given in
Eqn.~(\ref{rtp-eom}). The first equation shows that for super-horizon
modes when $-k\tau\rightarrow0$ one has
$|r_k|\rightarrow\infty$. Similarly $\phi_k\rightarrow\pi/2$ and
$\theta_k\rightarrow-\pi/2$ on super-horizon scales. We will later see
these parameters signify the probability distribution of the
wavefunctional in the phase space characterized by the Wigner function
\cite{Kiefer:2008ku, Kiefer:2006je, Kiefer:1998qe,
  Martin:2012pe}. While $r_k$ measures the excitation of the quantum
state, $\phi_k$ signifies the sharing of the excitation of the state
between the canonical variables \cite{Albrecht:1992kf}. In the
super-horizon limit we will see that in the standard inflationary
scenario the wavefunctional will become squeezed in the direction of
momentum conjugate to the field variable. Furthermore it can be argued
that such a squeezed Wigner function characterizes a classical
stochastic distribution of the field amplitude while the phase of the
field variable gets fixed. In other words, the quantum expectations can
be equivalently studied through averages over a classical stochastic
field \cite{Kiefer:2008ku, Kiefer:2006je, Kiefer:1998qe,
  Martin:2012pe}.

\subsubsection{Power spectrum}

The quantum fluctuations generated during inflation lead to quite a
few observational implications for the CMBR anisotropy spectrum we
observe today. One such important aspect of generic inflationary
scenarios is to predict a scale-invariant power spectrum of the
temperature fluctuations of the CMBR. To see that we first study the
power spectrum of the MS variable which is defined as the two-point
correlation function of these fluctuations:
\begin{eqnarray}
\langle0|\zeta(\tau,\mathbf{x})\zeta(\tau,\mathbf{x})|0\rangle=\int\frac{dk}{k}\mathcal{P}_{\zeta}(k),
\label{power-def}
\end{eqnarray}
where 
\begin{eqnarray}
\mathcal{P}_{\zeta}(k)=\frac{k^3}{2\pi^2}|f_k|^2.
\end{eqnarray}
Using Eqn.~(\ref{ms-R}) we see that the power spectrum of the comoving
curvature perturbation $\mathcal{R}$ is related to the power spectrum
of MS variable as
\begin{eqnarray}
\mathcal{P}_{\mathcal{R}}(k)=\left(\frac{\mathcal H}{a\varphi_0'}\right)^2\mathcal{P}_{\zeta}(k)=\frac{1}{2\epsilon a^2M^2_{\rm Pl}}\mathcal{P}_{\zeta}(k)\equiv A_Sk^{n_S-1},
\end{eqnarray}
where $A_S$ determines the amplitude of the power for wavenumber $k$
and the scalar spectral index $n_S$ provides the scale-dependence of
the power.

As has been mentioned before, the curvature perturbation $\mathcal{R}$
remains conserved on super-horizon scales and thus remains insensitive
to the complicated cosmological evolutions after inflation like
reheating. These primordial cosmological perturbations therefore can
be directly measured by measuring the perturbations in the temperature
$\left(\frac{\delta T}{T}\right)$ of the CMBR where these two kind of
perturbations are related to each other at the surface of
last-scattering as
\begin{eqnarray}
\left(\frac{\delta T}{T}\right)(\mathbf{e})=\frac15\mathcal{R}\left[\tau_{lss},-\mathbf{e}(\tau_{lss}-\tau_0)+\mathbf{x}_0\right],
\end{eqnarray}
where $\mathbf{e}$ is the direction in the sky where the temperature
fluctuation is measured, $\tau_{lss}$ and $\tau_0$ are the conformal
times at surface of last scattering and today respectively and
$\mathbf{x}_0$ is the present position from where the fluctuations are
measured. Hence the power spectrum of comoving curvature perturbations
is directly related to the two-point correlation function of the
temperature anisotropies in the CMBR which are accurately measured by
many present-day high precision observations such as WMAP
\cite{Hinshaw:2012fq} and PLANCK \cite{Ade:2013uln}. These
observations are all in agreement with a scale-invariant power
spectrum with $n_S\approx1$.

Analytically also we see that for the quasi-de Sitter case where
$a\sim-1/\tau$ and $|f_k|^2\sim 1/k^{3}\tau^{2}$ (using
Eqn.~(\ref{f-ksol})) for super-horizon modes, one obtains a
scale-invariant power spectrum for comoving curvature perturbations
which is in accordance with these observations and signifies that each
mode carries the same power during evolution.

\subsection{Schr\"{o}dinger Picture}

Now, we will try to correlate the Schr\"{o}dinger picture of evolution
of modes with the Heisenberg picture discussed in the previous section
following \cite{Polarski:1995jg}. In the Heisenberg picture the vacuum
$|0,\tau_0\rangle$ of the quantum state is defined at some time
$\tau_0$ as
\begin{eqnarray}
a_{\mathbf{k}}(\tau_0)|0,\tau_0\rangle_H=0.
\end{eqnarray}
One can see from Eqn.~(\ref{zetak}) and Eqn.~(\ref{pk}) that the vacuum
is an eigenstate of the operator $\hat\zeta_{\mathbf
  k}+i\Omega^{-1}_k\hat p_{\mathbf k}$:
\begin{eqnarray}
\left\{\hat\zeta_{\mathbf k}+i\Omega_k^{-1}(\tau)\hat p_{\mathbf k}\right\}|0,\tau_0\rangle_H=0,
\end{eqnarray}
where the time-dependent function $\Omega_k(\tau)$ is defined as
\begin{eqnarray}
\Omega_k=k\frac{u_k^*-v_k}{u_k^*+v_k}\equiv\frac{g_k^*}{f_k^*}=-i\frac{f_k^{*'}}{f_k^*}+i\frac{a'}{a}.
\label{Omega}
\end{eqnarray}

On the other hand, in the Schr\"{o}dinger picture the time-evolved
vacuum state satisfies the equation 
\begin{eqnarray}
\left\{\hat\zeta_{\mathbf k}(\tau_0)+i\Omega_k^{-1}(\tau)\hat p_{\mathbf k}(\tau_0)\right\}|0,\tau\rangle_S=0\,,
\end{eqnarray}
and the vacuum state $|0,\tau\rangle_S$ corresponds to a Gaussian
wave functional of the form 
\begin{eqnarray}
\Psi\left[\zeta_{\mathbf k}(\tau_0)\zeta_{-\mathbf k}(\tau_0)\right]=N_k(\tau)\exp\left(-\Omega_k\zeta_{\mathbf k}(\tau_0)\zeta_{-\mathbf k}(\tau_0)\right).
\label{functional}
\end{eqnarray}
Writing $\zeta_{\mathbf
  k}(\tau)$ in real and imaginary parts as 
\begin{eqnarray}
\zeta_{\mathbf  k}(\tau)=\frac{1}{\sqrt{2}}\left(\zeta_{\mathbf  k}^{\rm R}(\tau)+i\zeta^{\rm I}_{\mathbf  k}(\tau)\right),
\end{eqnarray}
allows us to write the real and imaginary Gaussian part of the
wave functional given in Eqn.~(\ref{functional}) as
\begin{eqnarray}
\Psi_{\mathbf k}^{\rm R,I}\left[\tau,\zeta^{\rm R,I}_{\mathbf k}\right]=\sqrt{N_k(\tau)}\exp\left(-\frac{\Omega_k(\tau)}{2}\left(\zeta_{\mathbf k}^{\rm R, I}\right)^2\right),
\label{psi-ri}
\end{eqnarray}
where the total wave function is
\begin{eqnarray}
\Psi\left[\zeta(\tau,\mathbf x)\right]=\prod_{\mathbf k}\Psi_{\mathbf k}\left[\zeta^{\rm R}_{\mathbf k}(\tau),\zeta^{\rm I}_{\mathbf k}(\tau)\right]=\prod_{\mathbf k}\Psi_{\mathbf k}^{\rm R}\left[\zeta^{\rm R}_{\mathbf k}(\tau)\right]\Psi_{\mathbf k}^{\rm I}\left[\zeta^{\rm I}_{\mathbf k}(\tau)\right]
\end{eqnarray}
As each mode $\mathbf{k}$ evolves independently and so do the real and
imaginary parts of their wavefunctional, for each mode $\mathbf{k}$
the real and imaginary parts of the wavefunctional $\Psi_{\mathbf
  k}^{\rm R,I}$ satisfy the functional Schr\"{o}dinger equation:
\begin{eqnarray}
i\frac{\partial \Psi_{\mathbf k}^{\rm
  R,I}}{\partial\tau}=\hat{\mathcal H}^{\rm R,I}_{\mathbf k}\Psi^{\rm R,I}_{\mathbf k},
\label{sch-eq}
\end{eqnarray}
where the Hamiltonian $\hat{\mathcal H}_{\mathbf k}\equiv
\hat{\mathcal H}^{\rm R}_{\mathbf k}+\hat{\mathcal H}^{\rm I}_{\mathbf
  k}$ is as given in Eqn.~(\ref{class-H}) but with operator $\hat p$
and $\hat \zeta$.

\subsubsection{Schr\"{o}dinger picture analysis with the equivalent action (without the boundary term) and the power spectrum}

We stated before that the two actions given in Eqn.~(\ref{action1})
and Eqn.~(\ref{action2}) are equivalent and analyzed the properties of
squeezing parameters considering the action given in
Eqn.~(\ref{action2}). The other action given in Eqn.~(\ref{action1})
has the advantage that the Hamiltonian of this action turns out to be
analogous to that for a harmonic oscillator:
\begin{eqnarray}
H\equiv\int d^3{\mathbf x}\,\mathcal{H}=\frac12\int d^3{\mathbf k}\left[p_{\mathbf k}p^*_{\mathbf k}+\left(k^2-\frac{z''}{z}\right)\zeta_{\mathbf k}\zeta^*_{\mathbf k}\right]
\label{class-H1}
\end{eqnarray}
where the conjugate momentum in Fourier space is 
\begin{eqnarray}
p_{\mathbf k}=\frac{\partial{\mathcal L}}{\partial \zeta_{\mathbf k}'}=\zeta^{*\prime}_{\mathbf k}.
\end{eqnarray}
This is very convenient for our analysis of the CSL mechanism in inflation
as the treatment of the harmonic oscillator of quantum mechanics within
CSL mechanism has been studied in the literature
\cite{Martin:2012pe}. Hence writing the Hamiltonian for the MS
variable modes in the form of a harmonic oscillator  helps us
incorporate CSL-like terms in its functional Schr\"{o}dinger equation.

If we consider the following action for the functional Schr\"{o}dinger
equation given in Eqn.~(\ref{sch-eq}) 
\begin{eqnarray}
\hat{\mathcal H}^{\rm R,I}_{\mathbf k}=-\frac12\frac{\partial^2}{\partial\left(\zeta^{\rm R,I}_{\mathbf k}\right)^2}+\frac12\omega^2\left(\zeta^{\rm R,I}_{\mathbf k}\right)^2,
\end{eqnarray}
where $\omega^2$ is given in Eqn.~(\ref{omega}), then one has to
determine the corresponding Gaussian functional $\Psi$ which would
satisfy Eqn.~(\ref{sch-eq}). Putting the real and imaginary part of
the wave functional into the functional Schr\"{o}dinger equation we get
\begin{eqnarray}
i\frac{N_k'}{N_k}=\Omega_k,\quad\quad \Omega_k'=-i\Omega_k^2+i\omega^2(\tau,k).
\label{N-Omega}
\end{eqnarray}
However, with this Hamiltonian one obtains $g_k^*=-if_k^{*'}$ which yields
\begin{eqnarray}
\Omega_k=k\frac{u_k^*-v_k}{u_k^*+v_k}\equiv\frac{g_k^*}{f_k^*}=-i\frac{f_k^{*'}}{f_k^*}.
\label{Omega1}
\end{eqnarray}
Clearly, putting this form of $\Omega_k$ in its evolution equation given in
Eqn.~(\ref{N-Omega}) yields the same equation of motion of $f_k$ as
given in Eqn.~(\ref{mode-eom}).

Now we can obtain the power spectrum in the Schr\"{o}dinger picture
following the steps in \cite{Martin:2012pe}. Knowing $\Omega_k$ one
can obtain the normalization of the wave functional as
\begin{eqnarray}
|N_k|=\left(\frac{\rm{Re}\,\Omega_k}{\pi}\right)^{\frac12}.
\end{eqnarray}
The two point correlation function of the Mukhanov-Sasaki variable is
defined as
\begin{eqnarray}
\langle \Psi|\zeta(\tau,\mathbf{x})\zeta(\tau,\mathbf{x})|\Psi\rangle=\int\prod_{\mathbf k}d\zeta^{\rm R}_{\mathbf k}d\zeta^{\rm I}_{\mathbf k}\,\Psi^*_{\mathbf k}(\zeta^{\rm R}_{\mathbf k},\zeta^{\rm I}_{\mathbf k})\,\zeta(\tau,\mathbf{x})\zeta(\tau,\mathbf{x})\,\Psi_{\mathbf k}(\zeta^{\rm R}_{\mathbf k},\zeta^{\rm I}_{\mathbf k}),
\end{eqnarray}
which yields \cite{Martin:2012pe}
\begin{eqnarray}
\langle \Psi|\zeta(\tau,\mathbf{x})\zeta(\tau,\mathbf{x})|\Psi\rangle=\frac{1}{(2\pi)^3}\int d{\mathbf k}\frac{1}{2 {\rm Re}\,\Omega_k}.
\end{eqnarray}
The real part of $\Omega_k$ can be easily obtained from
Eqn.~(\ref{Omega}) which turns out to be
\begin{eqnarray}
{\rm Re}\,\Omega_k=\frac{1}{2|f_k|^2},
\end{eqnarray}
using the Wronskian condition $W=i$. This gives the two-point
correlation function of Mukhanov-Sasaki variable as
\begin{eqnarray}
\langle \Psi|\zeta(\tau,\mathbf{x})\zeta(\tau,\mathbf{x})|\Psi\rangle=\frac{1}{(2\pi)^3}\int d{\mathbf k}\,|f_k|^2,
\end{eqnarray}
which yields the power spectrum of the Mukhanov-Sasaki variable
following Eqn.~(\ref{power-def}) as
\begin{eqnarray}
\mathcal{P}_{\zeta}(k)=\frac{k^3}{2\pi^2}|f_k|^2,
\end{eqnarray}
which is the same as obtained earlier in Heisenberg picture. In the
Schr\"{o}dinger picture as the wavefunctional is related to the
parameter $\Omega_k$ it would be convenient to define the
power-spectrum in terms of this parameter and thus writing
$|f_k|^2=1/2{\rm Re}\,\Omega_k$ the power spectrum of the
comoving curvature perturbations can be written as
\begin{eqnarray}
\mathcal{P}_{\mathcal{R}}(k)=\frac{k^3}{8\pi^2\epsilon M^2_{\rm Pl}}\frac{1}{a^2{\rm Re}\,\Omega_k}.
\label{power-R}
\end{eqnarray}
Thus in order to determine the nature of power spectrum of comoving
curvature perturbations we need to know the behavior of $a^2{\rm
  Re}\,\Omega_k$ in the super Hubble limit.
\subsection{Squeezed states and classicality through Wigner function}

\begin{figure*}[ht]
\begin{center}
\includegraphics[width=6.5cm, height=5cm]{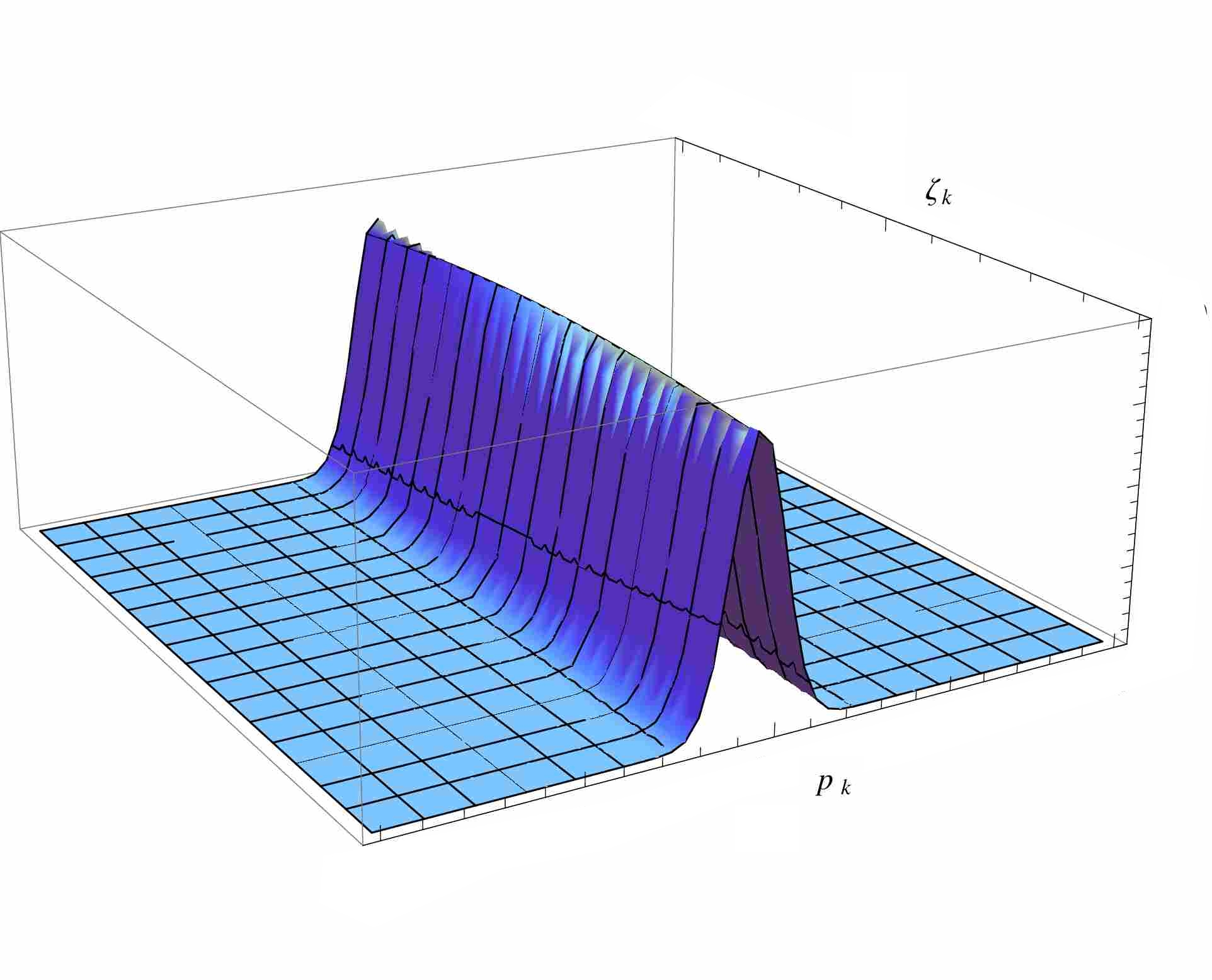}
\includegraphics[width=6.5cm, height=5cm]{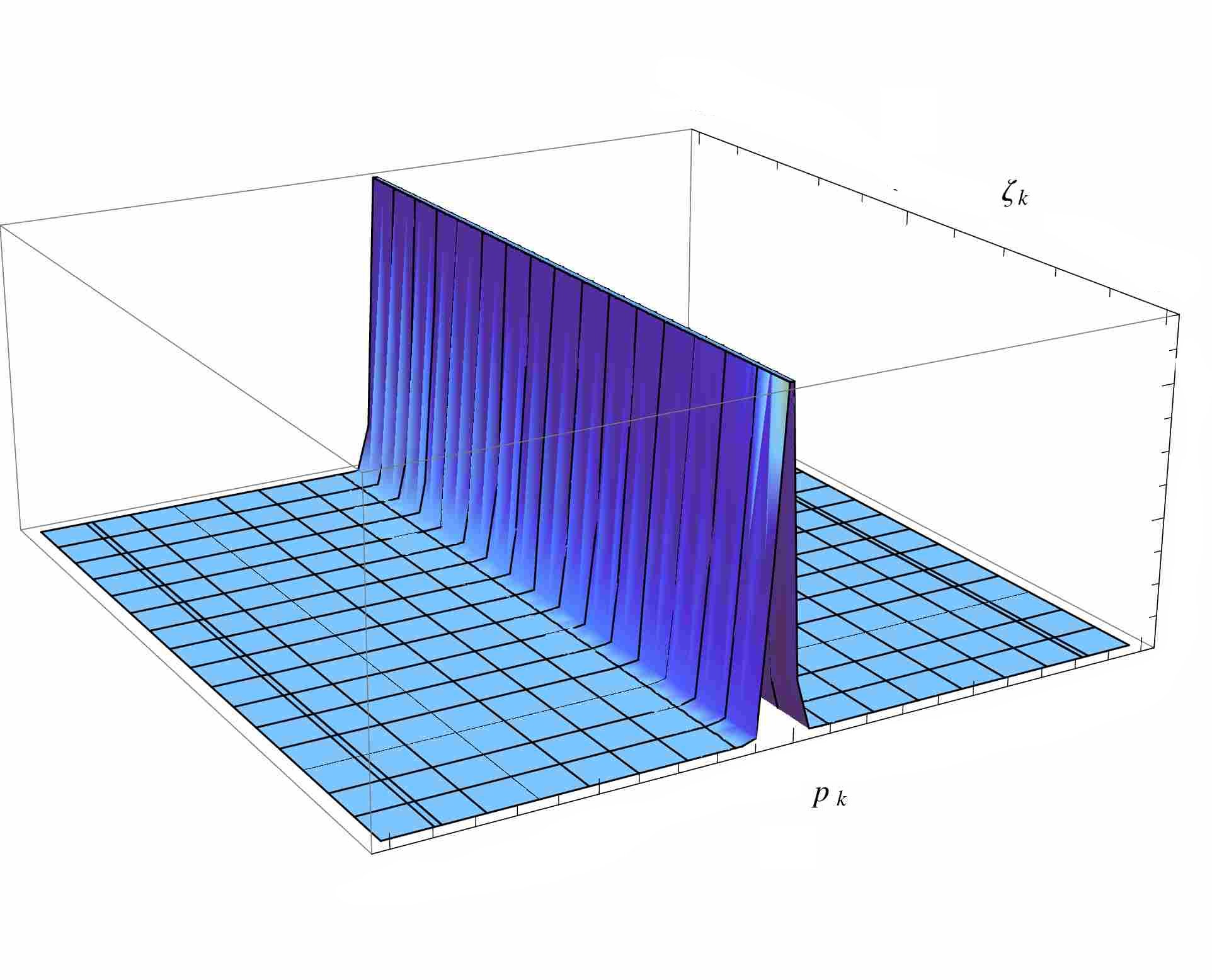}
\end{center}
\caption[]{Plot of Wigner function for $r_k=2.3$ and $r_k=3.5$ showing
  high squeezing of a particular mode in the direction of momentum
  canonical to field variable}
\label{wigner1}
\end{figure*}

We have earlier seen that on super-horizon scales the squeezing
parameter $|r_k|\rightarrow\infty$ and now we will determine the
correlation between this asymptotic limit of squeezing parameter $r_k$
and the classical nature of mode functions on super-horizon scales. In
general, to analyze the nature (quantum or classical) of mode
functions on super-horizon scales one determines the nature of the
Wigner function. In particular, a Wigner function recognizes the
correlation between position (in this case the field amplitude) and
momentum (canonical to the field in this case) and for a Gaussian
wavefunction it has positive values everywhere.  A positive definite
Wigner function can be interpreted as a classical probability
distribution of a quantum state under consideration in the phase
space. In quantum theory the Wigner function is defined by
\begin{eqnarray}
{\mathcal W}\left(\zeta^{\rm R}_{\mathbf k},\zeta^{\rm I}_{\mathbf k},p^{\rm R}_{\mathbf k},p^{\rm I}_{\mathbf k}\right)=\frac{1}{(2\pi)^2}\int dxdy\Psi^*\left(\zeta^{\rm R}_{\mathbf k}-\frac x2,\zeta^{\rm I}_{\mathbf k}-\frac y2\right)e^{-ip^{\rm R}_{\mathbf k}x-ip^{\rm I}_{\mathbf k}y}\Psi\left(\zeta^{\rm R}_{\mathbf k}+\frac x2,\zeta^{\rm I}_{\mathbf k}+\frac y2\right),
\end{eqnarray}
which then yields 
\begin{eqnarray}
{\mathcal W}\left(\zeta^{\rm R}_{\mathbf k},\zeta^{\rm I}_{\mathbf k},p^{\rm R}_{\mathbf k},p^{\rm I}_{\mathbf k}\right)&=&\frac{\Psi\Psi^*}{\pi{\rm Re}\,\Omega_k}\exp\left[-\frac{\left(p^{\rm R}_{\mathbf k}+{\rm Im}\,\Omega_k\zeta^{\rm R}_{\mathbf k}\right)^2}{{\rm Re}\,\Omega_k}\right]\exp\left[-\frac{\left(p^{\rm I}_{\mathbf k}+{\rm Im}\,\Omega_k\zeta^{\rm I}_{\mathbf k}\right)^2}{{\rm Re}\,\Omega_k}\right]\\
&=&\frac{1}{\pi^2}e^{-{\rm Re}\,\Omega_k\left(\zeta^{\rm R^2}_{\mathbf k}+\zeta^{\rm I^2}_{\mathbf k}\right)}e^{-\frac{\left(p^{\rm R}_{\mathbf k}+{\rm Im}\,\Omega_k\zeta^{\rm R}_{\mathbf k}\right)^2}{{\rm Re}\,\Omega_k}}e^{-\frac{\left(p^{\rm I}_{\mathbf k}+{\rm Im}\,\Omega_k\zeta^{\rm I}_{\mathbf k}\right)^2}{{\rm Re}\,\Omega_k}}.
\label{wigner-generic}
\end{eqnarray}
This shows that the Wigner function is a product of four Gaussians
with the first two Gaussians having a standard deviation of
$\sqrt{2/{\rm Re}\,\Omega_k}$ and zero mean while the last two
Gaussians with a standard deviation of $\sqrt{{\rm Re}\,\Omega_k/2}$
and mean of ${\rm Im}\,\Omega_k\zeta^{\rm R}_{\mathbf k}$ and ${\rm
  Im}\,\Omega_k\zeta^{\rm I}_{\mathbf k}$ respectively. To determine
the nature of the Wigner function for super-horizon modes we note that
\begin{eqnarray}
{\rm Re}\,\Omega_k&=&\frac{k}{\cosh(2r_k)+\cos(2\phi_k)\sinh(2r_k)},\nonumber\\
{\rm Im}\,\Omega_k&=&-\frac{k\sin(2\phi_k)\sinh(2r_k)}{\cosh(2r_k)+\cos(2\phi_k)\sinh(2r_k)}.
\end{eqnarray}
Hence on super-horizon scales when $|r_k|\rightarrow\infty$ one gets
${\rm Re}\,\Omega_k\rightarrow 0$. We also see from
Eqn.~(\ref{rtp-sol}) that $\phi_k\rightarrow\pi/2$ as
$-k\tau\rightarrow0$ on super-horizon scales. Thus ${\rm
  Im}\,\Omega_k\rightarrow0$ on super-horizon scales. In this strong
squeezing limit the last two Gaussians in the Wigner function would
become delta functions to yield
\begin{eqnarray}
{\mathcal W}\left(\zeta^{\rm R}_{\mathbf k},\zeta^{\rm I}_{\mathbf k},p^{\rm R}_{\mathbf k},p^{\rm I}_{\mathbf k}\right)\rightarrow \frac{{\rm Re}\,\Omega_k}{\pi}e^{-{\rm Re}\,\Omega_k\left(\zeta^{\rm R^2}_{\mathbf k}+\zeta^{\rm I^2}_{\mathbf k}\right)}\delta\left(p^{\rm R}_{\mathbf k}\right)\delta\left(p^{\rm I}_{\mathbf k}\right).
\end{eqnarray}
This indicates that the Wigner function will be highly squeezed in the
direction of $p_{\mathbf k}$ resulting in a cigar-like shape which has
been shown in Fig.~\ref{wigner1}.

This highly squeezed state successfully describes how the operator
expectations can be studied by considering averages over a classical
stochastic field \cite{Kiefer:2008ku, Kiefer:2006je, Kiefer:1998qe,
  Martin:2012pe}. However this description presumably excludes the
issue of localization of the initial perturbations in the field
variable as has been observed by several CMBR experiments. For more
discussions on this refer to \cite{Martin:2012pe}. Later we will show
the evolution of the Wigner function with CSL-like modification.
\section{CSL mechanism with a constant $\gamma$ and squeezing of modes during inflation}

\subsection{A Brief Overview of Continuous Spontaneous Localization}

CSL is a dynamical mechanism aimed at  answering the following conceptual issues in quantum theory: (1) Why are 
macroscopic objects not found in superposition of {\it position} states? (2) Why and how does the measurement process
break quantum mechanical state superposition? (3) What ``mass'', or, more accurately, how many nuclei,
should an object possess in order to qualify as a classical measuring apparatus? (4) Why do the outcomes of a quantum measurement obey the Born probability rule?

Following the pioneering work by Pearle \cite{Pearle:1976ka} and its
major improvement in the framework of the Ghirardi-Rimini-Weber (GRW)
model \cite{Ghirardi:1985mt}, the CSL model has been proposed as an
upgraded version of this model \cite{Pearle:1988uh,
  Ghirardi:1989cn}. CSL is designed to give a phenomenological answer
to the above four questions, in the case of non-relativistic quantum
mechanics. The basic physical idea behind dynamically induced
wave-function collapse is that spontaneous and random wave function
collapses occur all the time for all particles, whether isolated or
interacting and whether they are forming a microscopic, mesoscopic or
macroscopic system.

On a mathematical level, these ideas are implemented by modifying the Schr\"odinger equation while introducing extra
terms exhibiting the following properties: (1) non-linearity so as to yield the breakdown of superposition principle  at a
macroscopic level, (2) stochasticity so as to be able to explain the random outcomes of a measurement process and their
distribution according to the Born probability rule (stochasticity is moreover needed to avoid superluminal
communication), (3) allowance for an amplification mechanism according to which the new terms have negligible effects
on microscopic system dynamics but strong effects for large many-particle macroscopic systems in order to recover
their classical-like behavior.

To achieve this goal, a randomly fluctuating classical field, assumed to fill all of space, couples
to the quantum system number density operator to make the system collapse into its spatially localized eigenstates.
The collapse process being continuous in time, the dynamics can be described in terms of a single stochastic
differential equation containing features of both the standard Schr\"odinger evolution  and wave function collapse. An
outstanding open question is the origin of the random noise, or of the randomly fluctuating classical scalar field,
that induces the collapse. This is where an underlying fundamental theory is needed.

The physical content of the CSL model is nicely captured by a simple model known as QMUPL (Quantum Mechanics with Universal Position Localization), described by the following non-linear Schr\"odinger 
equation \cite{Diosi89}
\begin{equation} \label{eq:qmupl1}
d \psi_t  =  \left[ -\frac{i}{\hbar} H dt + \sqrt{\lambda} (q - \langle q \rangle_t) dW_t  
 - \frac{\lambda}{2} (q - \langle q \rangle_t)^2 dt \right] \psi_t
\end{equation}
Here, $\lambda$ is a new constant of nature which determines the strength of the coupling, and is assumed 
proportional to the mass $m$ of the particle: $\lambda/\lambda_{0} = m/m_{0}$. This is the amplification mechanism - the strength of the non-linear modification increases with increasing mass. The mass $m$ is expressed in units of the nucleon mass $m_0$. If one chooses for $\lambda_0$ the value $10^{-2}$ m$^{-2}$ s$^{-1}$ then the collapse of the wave-function can be explained consistently with all known experimental data. $W(t)$ is the standard Wiener process [Brownian motion] which encodes the stochastic aspect. As one can see, there are two non-linear, non-unitary terms in the modified equation, one proportional to $dW(t)$, and the other proportional to $dt$. Together, the two terms relate to each other in a specific way,  the latter being $(-1/2)$ times the former (since $dW^2 = dt$).  This `martingale' structure of the nonlinear equation preserves norm during evolution, despite being non-unitary, and is responsible for the emergence of the Born rule.

The stochastic term prevents a wave-packet from spreading indefinitely, and causes the width of the packet to reach a finite asymptotic value. For the chosen value of the constant $\lambda_{0}$ the spread is extremely small for a macroscopic object (so that it appears to be like a particle), and very large for a microscopic system such as an electron (so that it appears to be like a wave). Thus the modified quantum dynamics is able to describe in a universal and unified manner, both the wavy nature of microscopic systems, as well as the particle nature of macroscopic objects.

As an illustration of how these effects come about, consider for simplicity a free particle  $(H=p^2/2m)$ in the Gaussian state (the analysis can be generalized to other cases) \cite{Bassi2005, Bassi:2012bg}
\begin{equation} \label{gsol}
\psi_{t}(x) = \makebox{exp}\left[ - a_{t} (x -
\overline{x}_{t})^2 + i \overline{k}_{t}x + \gamma_{t}\right].
\end{equation}
By substituting this in the stochastic equation it can be proved that the spreads in position and momentum
\begin{equation}
\sigma_{q}(t)  \equiv  \frac{1}{2}\sqrt{\frac{1}{a_{t}^{\makebox{\tiny
R}}}};\qquad
\sigma_{p}(t)  \equiv  \hbar\,\sqrt{\frac{(a_{t}^{\makebox{\tiny R}})^2 +
(a_{t}^{\makebox{\tiny I}})^2}{a_{t}^{\makebox{\tiny R}}}},
\end{equation}
do not increase indefinitely but reach asymptotic values given by 
\begin{eqnarray} \label{aval1}
\sigma_{q}(\infty)& =& \sqrt{\frac{\hbar}{m\omega}} \simeq
\left( 10^{-15} \sqrt{\frac{\makebox{Kg}}{m}}\; \right)\,
\makebox{m}\, ,\nonumber\\
\sigma_{p}(\infty) &=& \sqrt{\frac{\hbar m\omega}{2}}
\simeq \left( 10^{-19} \sqrt{\frac{m}{\makebox{Kg}}}\,
\right)\, \frac{\makebox{Kg m}}{\makebox{s}},
\end{eqnarray}
such that:
$
\sigma_{q}(\infty)\, \sigma_{p}(\infty) \; = \;
{\hbar}/{\sqrt{2}}
$
which corresponds to almost the minimum allowed by Heisenberg's
uncertainty relations. Here, $\omega \; = \; 2\,\sqrt{{\hbar \lambda_{0}}/{m_{0}}} \; \simeq
\; 10^{-5} \; \makebox{s$^{-1}$}.$

The localization of the position state of macroscopic objects makes it straightforward to understand the collapse of the wave-function during a quantum measurement \cite{BassiSalvetti2007}. If a quantum system, say a two state quantum system, initially in a superposed state $(c_{+} |+\rangle + c_{-} |-\rangle)$, interacts with a measuring apparatus ${\cal A}$, the interaction causes the state to evolve to the entangled state
\begin{equation}
\label{eq:10}
c_{+}  |+ \rangle {\cal A}_{1} + c_{-} |- \rangle {\cal A}_{2}
\end{equation}
where ${\cal A}_{1}$ and ${\cal A}_{2}$ are respectively the pointer
states of the apparatus which correspond to the microscopic system
state $|+ \rangle$ and $|- \rangle$. Since, as we have seen above, the
stochastic dynamics prevents the apparatus (which is macroscopic) from
being simultaneously in the states ${\cal A}_{1}$ and ${\cal A}_{2}$,
this state `collapses' into either $c_{+} |+ \rangle {\cal A}_{1}$ or
$c_{-} |- \rangle {\cal A}_{2}$. The structure of the modified
equation ensures that the collapse probabilities obey the Born rule,
and the chosen value of the constant $\lambda_{0}$ ensures that the
collapse takes place sufficiently quickly.  It can be proved from
Eqn.~(\ref{eq:qmupl1}) that the initial state Eqn.~(\ref{eq:10})
evolves, at late times, to
\begin{equation}
  \label{eq:30}
  \psi_{t} = \frac{|+\rangle \otimes \phi_{+}
  + \epsilon_{t}|-\rangle \otimes \phi_{-}}{\sqrt{1+ \epsilon_t^2}}.
\end{equation}
The evolution of the stochastic quantity $\epsilon_t$ is determined dynamically by the stochastic equation: it either goes to $\epsilon_t \ll 1$, with a probability  $|c_{+}|^2$, or to $\epsilon_t \gg 1$, with a probability 
$|c_{-}|^2$. In the former case, one can say with great accuracy that the state vector has `collapsed' to the definite outcome $|+\rangle \otimes \phi_{+}$ with a probability $|c_{+}|^2$. Similarly, in the latter case one concludes that the state vector has collapsed to $|-\rangle \otimes \phi_{-}$ with a probability $|c_{-}|^2$.
This is how collapse during a quantum measurement is explained dynamically, and random outcomes over repeated measurements are shown to occur in accordance with the Born probability rule. The time-scale over which $\epsilon_t$ reaches its asymptotic value and the collapse occurs can also be computed dynamically. In the present example, for a pointer mass of 1 g, the collapse time turns out to be about $10^{-4}$ s.

The above QMUPL equation has been studied in quite some generality, and can be generalized to the multi-particle case as well, where it exhibits the all-important amplification property: the effective coupling strength parameter $\lambda_{eff}$ for the many-particle system scales as the total mass of the system, so that the localization effect is stronger for larger systems. The CSL model reproduces the above important properties [localization, dynamically induced collapse, amplification] of the QMUPL model while being able to deal with systems of indistinguishable particles. It is described by the following modified Schr\"odinger equation
\begin{eqnarray} \label{eq:csl-massa}
d\psi_t = \left[-\frac{i}{\hbar}Hdt + \frac{\sqrt{\gamma}}{m_{0}}\int d\mathbf{x} (M(\mathbf{x}) - \langle M(\mathbf{x}) \rangle_t)
dW_{t}(\mathbf{x}) -  \frac{\gamma}{2m_{0}^{2}} \int d\mathbf{x}\,
(M(\mathbf{x}) - \langle M(\mathbf{x}) \rangle_t)^2 dt\right] \psi_t\, .  \nonumber\\
\end{eqnarray}
The linear part is governed by $H$ - the standard quantum Hamiltonian of the system, and like in QMUPL, the other two
terms induce the collapse of the wave function in space. The mass $m_0$ is a reference mass, which as before is taken
equal to that of a nucleon. Analogous to $\lambda$, the parameter $\gamma$ is a {\it mass-proportional} [and hence possesses the amplification property] positive coupling
constant which sets the strength of the collapse process, while $M({\bf x})$ is
a smeared {\it mass density} operator:
\begin{eqnarray}
M(\mathbf{x})
& = & \underset{j}{\sum}m_{j}N_{j}(\mathbf{x}), \nonumber\label{eq:dsfjdhz}\\
N_{j}(\mathbf{x})
& = & \int d\mathbf{y}g(\mathbf{y-x})
\psi_{j}^{\dagger}(\mathbf{y})\psi_{j}(\mathbf{y}), \qquad
\end{eqnarray}
$\psi_{j}^{\dagger}(\mathbf{y})$,
$\psi_{j}(\mathbf{y})$ being, respectively, the creation and
annihilation operators of a particle of type $j$ in the space point
$\mathbf{y}$. The smearing function $g({\bf x})$ is taken equal to
\begin{equation} \label{eq:nnbnm}
g(\mathbf{x}) \; = \; \frac{1}{\left(\sqrt{2\pi}r_{c}\right)^{3}}\;
e^{-\mathbf{x}^{2}/2r_{C}^{2}},
\end{equation}
where $r_C$ is the second new phenomenological constant of the model. 

The proof for the dynamical collapse of the wave-function and the
emergence of the Born probability rule in the CSL model can be found
for instance in Section III.A.7 of \cite{Bassi:2012bg}. The basic idea
is that stochastic fluctuations drive to zero the variance $V\equiv
\langle (A-\langle A \rangle )^2\rangle$ of the operator $A\equiv \int
d\mathbf{x}\, M(\mathbf{x})$. As a consequence, the system is driven
to one of the eigenstates of the measured observable. The
norm-preserving martingale structure of the CSL equation ensures that
the stochastic expectation of the projection operator is preserved:
this coincides with the square of the amplitude in a given state,
initially, and finally with the probability to result in that
particular state, thus establishing the Born rule.

The CSL model is reviewed in some detail in \cite{Bassi:2012bg}. An
attractive feature of the model is that it is experimentally
falsifiable with currently amenable technology, as its predictions
depart from those of quantum theory in the mesoscopic regime. The
model is being subjected to rigorous experimental tests which include
molecular interferometry, optomechanics, and bounds on its fundamental
parameters from astrophysical and cosmological observations
\cite{Bassi:2012bg}.

A novel feature of the model is that it predicts a very tiny violation of energy and momentum conservation, because of the presence of the stochastic process. While on the one hand this violation is too small to contradict known physics, it has also been suggested that such a violation induces anomalous Brownian motion, which may be detectable in laboratory experiments with mesoscopic systems 
\cite{CollettPearle03}.

It must be noted that CSL is a non-relativistic model, by construction. The collapse of the wave-function is an instantaneous process. 
While it is well understood that instantaneous collapse cannot be used for superluminal signaling, it is nonetheless an `action at a distance' feature,
 which is not in accord with special relativity. A relativistic version of CSL would be highly desirable, but has not been achieved yet \cite{Bassi:2012bg}
 and it has even been suggested that if verified, the CSL formalism might possibly hint at the need for a drastic revision of some basic concepts relating quantum mechanics and special
 relativity.

\subsection{Application of a CSL-like Mechanism to Inflation}

The first obstacle in applying the CSL mechanism in the inflationary
paradigm to explain the classical transition of quantum mode
fluctuations is that a relativistic Quantum Field Theory version of
CSL model is yet to be developed and attempts to construct a viable
relativistic field theoretic model of CSL face numerous problems
including irremovable divergences \cite{Pearle:2005rc, Pearle:1976ka,
  Ghirardi:1985mt, Pearle:1988uh, Ghirardi:1989cn, Bassi:2003gd,
  Weinberg:2011jg}. In spite of the lack of a proper quantum field
theoretic version of CSL, Martin {\it et al.} in \cite{Martin:2012pe}
made an attempt to modify the functional Schr\"{o}dinger equation of
mode functions in Fourier space by adding `CSL-like' stochastic terms
with spontaneous localization on the $\hat\zeta_{\mathbf k}$
eigenmanifolds. The presence of stochasticity in the functional
Schr\"{o}dinger equation can be motivated from the fact that the
non-relativistic limit of such a field theory should reproduce the
known CSL stochastic evolution. In the non-relativistic CSL mechanism
the stochastic evolution is `position-driven' \cite{Bassi:2003gd,
  Bassi:2012bg} whereas in \cite{Martin:2012pe} the addition of
`CSL-like' stochastic terms has been done in the Fourier basis to
study the inflationary quantum fluctuations. Such a departure from the
standard CSL formalism for the inflationary theories can be justified
knowing that the presence of primordial non-Gaussianities are
negligible so that different Fourier modes evolve
independently. Having modified the functional Schr\"{o}dinger equation
in the field basis one would be lead to convolution of field modes in
the Fourier space which will render the Gaussian nature of primordial
fluctuations. Also, in such a modification the `CSL-like' parameter
$\gamma$ turns out to be of mass dimension 2 which is quite different
from its non-relativistic version.  In such a case, the bounds on the
CSL parameter $\gamma$ coming from the inflationary scenario should,
strictly speaking, are not to be compared with those coming from other
quantum mechanical systems.

Here, we will also follow the same method of modifying the functional
Schr\"{o}dinger equation developed by Martin {\it et al.} for
inflationary dynamics.  We will consider the CSL evolution which is
driven by the Mukhanov-Sasaki variable. It is important to note here that though
Martin {\it et al.} added `CSL-like terms' to the functional
Schr\"{o}dinger equation, they dealt with a constant CSL-like
parameter $\gamma$  in MS variable driven model and in doing so the formalism lacks 
the aforementioned scale-dependent
amplification mechanism. Also an inflationary CSL mechanism with a
constant $\gamma$ term yields a localization in the conjugate momentum
direction as in a generic case of inflationary scenario. Thus CSL
mechanism with a constant $\gamma$ does not seem to have an advantage
over the generic inflationary scenario. In this section we will
discuss the inflationary CSL mechanism with constant $\gamma$ part and
investigate the features of squeezing in this scenario. In the next
section, we will present a CSL type modification which incorporates
the amplification mechanism, and also explains the quantum to
classical transition.

The modified functional Schr\"{o}dinger equation with a constant
CSL-like parameter $\gamma$ for Mukhanov-Sasaki variable is written as
\cite{Martin:2012pe}
\begin{eqnarray}
d\Psi_{\mathbf k}^{\rm R,I}=\left[-i\hat{\mathcal{H}}_{\mathbf k}^{\rm R,I}d\tau+\sqrt{\gamma}\left(\hat\zeta_{\mathbf k}^{\rm R,I}-\left\langle\hat\zeta_{\mathbf k}^{\rm R,I}\right\rangle\right)dW_\tau-\frac{\gamma}{2}\left(\hat\zeta_{\mathbf k}^{\rm R,I}-\left\langle\hat\zeta_{\mathbf k}^{\rm R,I}\right\rangle\right)^2d\tau\right],
\label{func-sch-eq}
\end{eqnarray}
where the stochastic behavior due to CSL mechanism is encoded in the
Wiener process $W_\tau$. We observe the formal similarity of this
equation with the original CSL Eqn. (\ref{eq:csl-massa}). The most
general stochastic wave-functional which satisfies this stochastic
functional Schr\"{o}dinger equation can be written as
\begin{eqnarray}
\Psi^{\rm R,I}_{\mathbf k}\left(\tau,\zeta_{\mathbf k}^{\rm R,I} \right)&=&\left|\sqrt{N_k(\tau)}\right|\exp\left\{-\frac{{\rm Re}\,\Omega_k(\tau)}{2}\left[\zeta_{\mathbf k}^{\rm R,I}-\bar\zeta_{\mathbf k}^{\rm R,I}(\tau)\right]^2+i\sigma_{\mathbf k}^{\rm R,I}(\tau)+i\chi_{\mathbf k}^{\rm R,I}(\tau)\zeta_{\mathbf k}^{\rm R,I}\right.\nonumber\\
&&\left.\quad\quad\quad\quad\quad\quad\quad-i\frac{{\rm Im}\,\Omega_k(\tau)}{2}\left(\zeta_{\mathbf k}^{\rm R,I}\right)^2\right\},
\end{eqnarray}
where $\bar\zeta_{\mathbf k}^{\rm R,I}$, $\sigma_{\mathbf k}^{\rm
  R,I}$ and $\chi_{\mathbf k}^{\rm R,I}$ are real numbers. If we put
$\bar\zeta_{\mathbf k}^{\rm R,I}$, $\sigma_{\mathbf k}^{\rm R,I}$ and
$\chi_{\mathbf k}^{\rm R,I}$ to be zero then the wave function matches
with that given in Eqn.~(\ref{psi-ri}). A set of differential equations
followed by the functions parameterizing the above functional Gaussian
state is given in \cite{Martin:2012pe} which we quote here for
completeness:
\begin{eqnarray}
\label{eq:evol:N}
\frac{\vert N_k\vert ^\prime}{\vert N_k \vert}&=&
\frac12\frac{\left({\rm Re}\,\Omega_k\right)^\prime}{{\rm Re}\, \Omega_k}={\rm Im}\,\Omega_k+\frac{\gamma}{4{\rm Re}\,  \Omega_k},\nonumber\\
\label{eq:evol:Omega}
\left({\rm Re}\,  \Omega_k\right)^\prime &=&2\gamma+2\left({\rm Re}\, \Omega_k\right)\left({\rm Im}\,  \Omega_k\right),\nonumber\\
\label{eq:evol:h}
\left({\rm Im}\,  \Omega_k\right)^\prime &=&-\left({\rm Re}\,\Omega_k\right)^2+\left({\rm Im}\,\Omega_k\right)^2+\omega^2\left(\tau,k\right),\nonumber\\
\label{eq:evol:barnu}
\left(\bar{\zeta}_{\mathbf{k}}^\mathrm{R,I}\right)^\prime&=&\chi_{\mathbf{k}}^\mathrm{R,I}+\frac{\sqrt{\gamma}}{{\rm Re}\,\Omega_k}\frac{d W_\tau}{d \tau}-\left({\rm Im}\,\Omega_k\right)\bar{\zeta}_{\mathbf k}^\mathrm{R,I},\nonumber\\
\label{eq:evol:f}
\left(\sigma_{\mathbf{k}}^\mathrm{R,I}\right)^\prime&=&
-\frac12{\rm Re}\,\Omega_k+\frac12\left({\rm Re}\,\Omega_k\right)^2\left(\bar{\zeta}_{\mathbf{k}}^\mathrm{R,I}\right)^2-\frac12\left(\chi_{\mathbf{k}}^\mathrm{R,I}\right)^2, \nonumber\\
\label{eq:evol:g}
\left(\chi_{\mathbf{k}}^\mathrm{R,I}\right)^\prime&=&-\left({\rm Re}\,\Omega_k\right)^2\bar{\zeta}_{\mathbf{k}}^\mathrm{R,I}+\chi_{\mathbf{k}}^\mathrm{R,I}\left({\rm Im}\,\Omega_k\right).
\end{eqnarray}
We see from the above set of equations that the evolution of $N_k$ and
$\Omega_k$ does not depend upon the other parameters of the
wave functional. Also we have seen before that $N_k$ and $\Omega_k$ are
the two parameters required to determine the Wigner function. From the
first equation of the set of equations given above we get
\begin{eqnarray}
|N_k|=\left(\frac{\rm{Re}\,\Omega_k}{\pi}\right)^{\frac12}
\end{eqnarray}
and combining the second and the third equations one gets:
\begin{eqnarray}
\Omega_k'=-i\Omega_k^2+i\omega^2(\tau,k)+2\gamma=-i\Omega_k^2+i\tilde\omega^2(\tau,k),
\label{Omega-eom}
\end{eqnarray}
where
$\tilde\omega^2\equiv\omega^2-2i\gamma=k^2-2i\gamma (a''/a)$. We
define:
\begin{eqnarray}
\kappa^2=k^2-2i\gamma,
\end{eqnarray}
which allows one to write $\tilde\omega^2=\kappa^2-a''/a$, the same
form of $\omega^2$ as given in Eqn.~(\ref{omega}). By analogy with the
discussion provided in the previous section, we can now show that the
function $f_k$ too would satisfy the same equation of motion as given
in Eqn.~(\ref{mode-eom}) but now with frequency $\tilde\omega$:
\begin{eqnarray}
f_k''+\left(k^2-2i\gamma-\frac{a''}{a}\right)f_k=0.
\label{mode-eom-1}
\end{eqnarray}
Before we proceed with this analysis, it is very important to point
out one important difference between Schr\"{o}dinger picture analysis
of standard inflationary scenario and that of the one with modified
`CSL-like' terms. We would like to emphasize the point that once the
inflationary dynamics is modified with `CSL-like' terms we no longer
have the corresponding Heisenberg picture as we yet do not know the
Lagrangian formulation of CSL dynamics. In such a case we will analyze
all the relevant observable quantities (such as power spectrum) in the
Schr\"{o}dinger picture which can be determined in terms of
$\Omega_k$. Also, in Schr\"{o}dinger picture $f_k$ is rather a
parameter, which would help one to determine the functional form of
$\Omega_k$, than the mode function in Heisenberg picture. Unless a
viable Lagrangian formulation of CSL dynamics is achieved, it would be
difficult to relate the parameter $f_k$ in Schr\"{o}dinger picture
with mode functions in Heisenberg picture. Also, we will keep writing
$f_k$ as we have done in the standard inflationary scenario but
keeping in mind that with CSL-modified dynamics this parameter can not
be considered as mode function any longer and would be treated only as
a parameter with no a prior observational significance.

In \cite{Martin:2012pe} an exact solution of the above equation is
obtained which turns out to be Bessel functions whose asymptotic
limits are known. Therefore knowing the asymptotic behavior of $f_k$
in the super-horizon limit one can construct the power spectrum in
terms of ${\rm Re}\,\Omega_k$ which turns out to be scale-dependent for
large modes.  However, we will be interested in the case where the
`CSL-like' parameter would be scale-dependent. In those cases exact
solutions of mode functions will not be available. Moreover, we would
like to investigate the nature of the Wigner function to see the
effects of CSL modification for classicalization of modes. Therefore
we will study the evolution of modes in terms of the squeezing parameters
which we describe below.

First we will derive the evolution equations of squeezing parameters
with `CSL-like' modifications. Taking the complex conjugate of the
above equation it now shows that $f_k^*$ is not a solution of the same
equation and for defining the Wronskian in this case we will need two
independent solutions of Eqn.~(\ref{mode-eom-1}). Therefore, we define
an operation :
\begin{eqnarray}
\bar f_k\equiv f_k^*\,\,({\rm with}\,\,\gamma\rightarrow-\gamma),
\end{eqnarray}
to see that under such an operation Eqn.~(\ref{mode-eom-1}) becomes
\begin{eqnarray}
\bar f_k''+\left(k^2-2i\gamma-\frac{a''}{a}\right)\bar f_k=0,
\end{eqnarray}
which shows that $f_k$ and $\bar f_k$ satisfies the same equation of
motion. Similarly, we define $\Omega_k$ for this case as 
\begin{eqnarray}
\Omega_k=-i\frac{\bar f_k'}{\bar f_k}
\end{eqnarray}
and putting it back in Eqn.~(\ref{Omega-eom}) gives the same equation
of motion of $\bar f_k$ (and so for $f_k$) as written above. 

One then can straightforwardly write all the other equations related
to the squeezing of modes discussed before by replacing $x^*$ with
$\bar x$ (for a complex variable $x$) and $k$ with $\kappa$. For
completeness we write the necessary equations here once again. First
of all, the Wronskian for this system would be
\begin{eqnarray}
W=f_k\bar f_k'-\bar f_kf_k'=i.
\end{eqnarray}
The function $f_k$ is now related to $u_k$ and $v_k$ as
\begin{eqnarray}
f_k=\frac{u_k+\bar v_k}{\sqrt{2\kappa}}
\label{f-uv1}
\end{eqnarray}
and the Wronskian will yield $u_k\bar u_k-v_k\bar v_k=1$. This would
allow one to parameterize $u_k$ and $v_k$ in terms of squeezing
parameters as we did before :
\begin{eqnarray}
u_k(\tau)&=&e^{-i\theta_k(\tau)}\cosh r_k(\tau),\nonumber\\
v_k(\tau)&=&e^{i\theta_k(\tau)+2i\phi_k(\tau)}\sinh r_k(\tau),
\label{uv-rtp}
\end{eqnarray}
where now $r_k$, $\phi_k$ and $\theta_k$ are complex quantities. We
assume that $\bar r_k=r_k$, $\bar \phi_k=\phi_k$ and $\bar
\theta_k=\theta_k$. The evolution equations followed by $u_k$ and
$v_k$ are
\begin{eqnarray}
u_k'&=&-i\kappa u_k+\frac{a'}{a}\bar v_k\,,\nonumber\\
v_k'&=&-i\kappa v_k+\frac{a'}{a}\bar u_k\,,
\end{eqnarray}
which allows one to write $\Omega_k$ as
\begin{eqnarray}
\Omega_k\equiv-i\frac{\bar f_k'}{\bar f_k}=\kappa\frac{\bar u_k-v_k}{\bar u_k+v_k}.
\label{Omega-const}
\end{eqnarray}
One can also derive the evolution equations of the squeezing
parameters as
\begin{eqnarray}
r_k'&=&\frac{a'}{a}\cos2\phi_k,\nonumber\\
\phi_k'&=&-\kappa-\frac{a'}{a}\sin 2\phi_k\coth 2r_k,\nonumber\\
\theta_k'&=&\kappa+\frac{a'}{a}\sin 2\phi_k\tanh r_k.
\label{rtp-eom-1}
\end{eqnarray}
With constant $\gamma$ the above equations have exact solutions (same
as given in Eqn.~(\ref{rtp-sol}) with $k$ replaced by $\kappa$) with
$a(\tau)\sim-1/\tau$. But we would like to introduce an approximation
scheme to solve for super-horizon modes because it will help us in
further discussion where the CSL-like parameter $\gamma$ would become
time-dependent and thus exact solutions of evolution equations of
squeezing parameters will not be available.
\subsection{Approximate solutions of squeezing parameters}

We would require to solve for $r_k$ and $\phi_k$ only as these are
required to determine ${\rm Re}\,\Omega_k$ which yields the nature of
Wigner function of the system. Let us first assume that the real part
of $r_k$ (we have written $r_k=r_k^{\rm R}+ir_k^{\rm I}$) becomes
$r_k^{\rm R}\rightarrow -\infty$ as $-k\tau\rightarrow0$ (this
assumption will be verified later). Under this assumption
$\coth(2r_k)\rightarrow -1$ in superhorizon limit. This simplifies the
evolution equations of $r_k$ and $\phi_k$ in superhorizon limit which
become
\begin{eqnarray}
r_k'&=&-\frac1\tau\cos2\phi_k,\nonumber\\
\phi_k'&=&-\kappa-\frac1\tau\sin 2\phi_k.
\end{eqnarray}
Defining $y=-\kappa\tau$ the above two equations can be rewritten as
\begin{eqnarray}
\frac{dr_k}{dy}&=&\frac1y\cos2\phi_k,\nonumber\\
\frac{d\phi_k}{dy}&=&1-\frac1y\sin 2\phi_k.
\end{eqnarray}
To solve for $\phi_k$ we use the transformation
$\tan\phi_k=\mathcal{F}_k(y)$ to yield 
\begin{eqnarray}
\frac{\partial\mathcal{F}_k}{\partial y}-\mathcal{F}_k^2+\frac2y\mathcal{F}_k=1,
\end{eqnarray}
which has a solution
\begin{eqnarray}
\tan\phi_k\equiv\mathcal{F}_k(y)=\frac1y+\tan y.
\end{eqnarray}
Putting this in the evolution equation of $r_k$ gives
\begin{eqnarray}
\frac{dr_k}{dy}=-\frac1y\left(\frac{y^2-1}{y^2+1}\right),
\end{eqnarray}
which yields a solution for $r_k$ as 
\begin{eqnarray}
r_k=\ln(y)-\ln(1+y^2).
\end{eqnarray}
Hence, at superhorizon limit we get
\begin{eqnarray}
r_k&\rightarrow&\ln y\,,\nonumber\\
\tan\phi_k&\rightarrow&\frac1y.
\end{eqnarray}
Writing explicitly the real and imaginary parts of $\kappa$, $r_k$
and $\phi_k$ :
\begin{eqnarray}
\kappa^{\rm R} &=& k\left(1+\frac{4 \gamma^2}{k^4} \right)^{\frac14}\cos\left(\frac12\tan^{-1}\left(\frac{2\gamma}{k^2}\right)\right)\,,\nonumber\\
\kappa^{\rm I} &=& -k\left(1+\frac{4 \gamma^2}{k^4} \right)^{\frac14}\sin\left(\frac12\tan^{-1}\left(\frac{2\gamma}{k^2}\right)\right)\,,\nonumber\\
r_k^{\rm R}&=&  \ln\left(-k\tau\left(1+\frac{4 \gamma^2}{k^4} \right)^{\frac14}\right),\nonumber\\
r_k^{\rm I} &=&-\frac12\tan^{-1}\left(\frac{2\gamma}{k^2} \right),\nonumber\\
\tan(\phi_k^{\rm R})&=&\frac{1}{(-k\tau)}\left(1+\frac{4\gamma^2}{k^4}\right)^{-\frac14}\sec\left(\frac12\tan^{-1}\left(\frac{2\gamma}{k^2}\right)\right),\nonumber\\
\tanh\left(\phi_k^{\rm I}\right)&=&(-k\tau)\left(1+\frac{4 \gamma^2}{k^4}\right)^{\frac14}\sin\left(\frac12\tan^{-1}\left(\frac{2\gamma}{k^2}\right)\right),
\label{real-im-parts}
\end{eqnarray}
we see that in the superhorizon limit when $-k\tau\rightarrow0$ one
gets $r_k^{\rm R}\rightarrow-\infty$ (which we have assumed earlier)
and $\phi_k^{\rm R}\rightarrow\pi/2$ as we have in the generic
inflationary case. We also have $\phi_k^{\rm I}\rightarrow0$ in this
limit. These solutions also verify our assumptions $\bar r_k=r_k$,
$\bar \phi_k=\phi_k$ and $\bar \theta_k=\theta_k$ while also tally
with the asymptotic limits of the exact solution.
\subsection{Classicality and Power Spectrum with constant $\gamma$}

We have discussed before that the nature of classicality is determined
by the nature of Wigner function. We also noticed that ${\rm
  Re}\,\Omega_k$ plays an important role in determining in which
direction the modes will be squeezed if macro-objectification of the
modes has to  occur in the evolution. Writing ${\rm Re}\,\Omega_k$ from
Eqn.~(\ref{Omega-const}) explicitly in terms of real and imaginary
parts of $\kappa$, $r_k$ and $\phi_k$ we get 
\begin{eqnarray}
{\rm
  Re}\,\Omega_k=\frac{\kappa^{\rm R}\left[(1+e^{4\phi_k^{\rm I}})\cos{2r_k^{\rm I}}+(e^{4\phi_k^{\rm I}}-1)\cosh{2r_k^{\rm R}}\right]+2e^{2\phi_k^{\rm I}}\kappa^{\rm I}\left[\cos{2\phi_k^{\rm R}}\sin{2r_k^{\rm I}}+
  \sin{2\phi_k^{\rm  R}}\sinh{2r_k^{\rm R}}\right]}{(e^{4\phi_k^{\rm I}}-1)\cos{2r_k^{\rm I}}+(1+e^{4\phi_k^{\rm I}})\cosh{2r_k^{\rm R}}-2e^{2\phi_k^{\rm I}}(\sin{2\phi_k^{\rm R}}\sin{2r_k^{\rm I}}-
  \cos{2\phi_k^{\rm R}}\sinh{2r_k^{\rm R}})}.\nonumber\\
\label{re-omega-k}
\end{eqnarray} 

We note at this point that the real and imaginary parts of $\kappa$,
$r_k$ and $\phi_k$ depend upon the wavenumber $k$ and the CSL
parameter $\gamma$ as can be seen from Eqn.~(\ref{real-im-parts}). The
ratio $\gamma/k^2$ thus naturally sets a scale in this theory. First
we will analyze the nature of the largest modes for which the
wavenumber number $k$ is very small and thus $2\gamma/k^2\gg1$. For
these modes we have
\begin{eqnarray}
\begin{array}{ccc}
\kappa^{\rm R}\sim \sqrt{\gamma}, & & \kappa^{\rm I}\sim -\sqrt{\gamma},\\
r_k^{\rm R}\sim \ln(-\sqrt{2\gamma}\tau), & & r_k^{\rm I}\sim -\frac\pi4,\\
\phi_k^{\rm R}\sim \tan^{-1}\left(-\frac{1}{\sqrt{\gamma}\tau}\right), & & \phi_k^{\rm I}\sim -\sqrt{\gamma}\tau,
\end{array}
\end{eqnarray}
which yield 
\begin{eqnarray}
{\rm Re}\,\Omega_k\approx \frac{2\gamma}{k}(-k\tau),
\label{large}
\end{eqnarray}
up to leading order in $-k\tau$. This shows that ${\rm
  Re}\,\Omega_k\rightarrow0$ as $-k\tau\rightarrow0$ as it happens in
a generic inflationary case. Thus the squeezing of largest modes will
be in the direction of momentum of the field $\zeta$ as before. This
is precisely the regime where one would have expected `CSL-like'
modification to dominate the evolution and cause an effective collapse
in the field basis. However we see from the Wigner function analysis
that the modes remain squeezed in the momentum direction for constant
$\gamma$ case. 
\begin{figure*}[ht]
\begin{center}
\includegraphics[width=8cm, height=5.5cm]{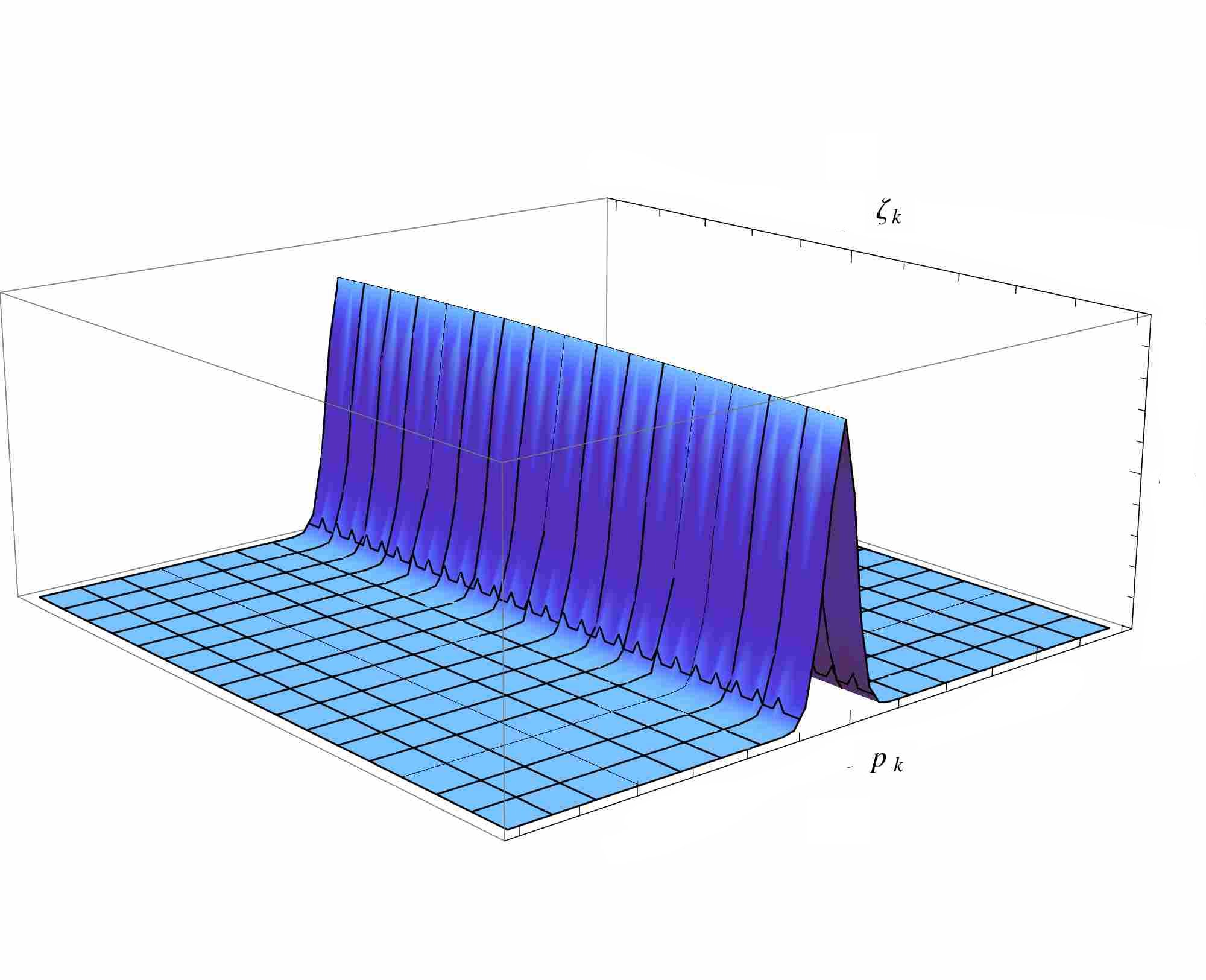}
\end{center}
\caption[]{Plot of Wigner function in the case with constant $\gamma$
  for $|r_k|=6.6$ and $\gamma=100$ showing high squeezing of a
  particular mode in the direction of momentum canonical to field
  variable}
\label{wigner2}
\end{figure*}

For the smaller modes for which $2\gamma/k^2\ll1$ one has
\begin{eqnarray}
\begin{array}{ccc}
\kappa^{\rm R}\sim k, & & \kappa^{\rm I}\sim 0,\\
r_k^{\rm R}\sim \ln(-k\tau), & & r_k^{\rm I}\sim 0,\\
\phi_k^{\rm R}\sim \tan^{-1}\left(-\frac{1}{k\tau}\right), & & \phi_k^{\rm I}\sim 0,
\end{array}
\end{eqnarray}
which yields 
\begin{eqnarray}
{\rm Re}\,\Omega_k\approx2k(-k\tau)^2,
\label{small}
\end{eqnarray}
which also tends to zero as $-k\tau\rightarrow0$ yielding squeezing in
the direction of momentum for the smaller modes. In Fig.~\ref{wigner2}
we can see that despite the presence of a CSL correction with a
constant $\gamma$ the Wigner function shows that the wavefunctional
collapse has not taken place in the field eigenbasis and hence this
scenario fails to explain the macro-objectification of the
inflationary quantum fluctuations.

Now let us see the nature of power spectrum due to the effects of CSL
mechanism. From Eqn.~(\ref{power-R}) we see that the scale-dependence
of the power arises from the factor $a^2{\rm Re}\,\Omega_k$. Also from the
above discussion we see that ${\rm Re}\,\Omega_k$ depends upon time
$\tau$ for both long and short modes making the power spectrum
time-dependent which is quite contrary to a generic inflationary
scenario. This phenomenon has also been observed in
\cite{Martin:2012pe} where the authors  have evaluated the power
at the end of inflation. Following the same method as in
\cite{Martin:2012pe} we rewrite $-k\tau$ as
\begin{eqnarray}
-k\tau=\frac{k}{k_0}e^{-\Delta N},
\label{end-inf}
\end{eqnarray}
where $k_0$ is the comoving wave number of the mode which is at the
horizon today $k_0=a_0H_0$ and $\Delta N$ is the number of e-foldings
the mode $k$ has spent outside the Hubble radius during inflation and
thus $\Delta N\approx 50-60$ for observationally relevant modes.  It
must be noted that it is not obvious in such a scenario whether the
power spectrum calculated at the end of inflation would be the same at
recombination. But as these modes are superhorizon at the end of
inflation, one can expect the causal physics after inflation to not
affect these modes considerably.

Hence for shorter modes we have $a^2{\rm Re}\,\Omega_k=2k^3$ which
yields the power spectrum using Eqn.~(\ref{power-R}) as
\begin{eqnarray}
\mathcal{P}_{\mathcal{R}}(k)=\frac{1}{16\pi^2\epsilon M^2_{\rm Pl}},
\end{eqnarray}
which is a scale-invariant power spectrum. On the other hand for
longer modes one gets $a^2{\rm Re}\,\Omega_k=2\gamma k_0e^{\Delta N}$
which yields a power spectrum 
\begin{eqnarray}
\mathcal{P}_{\mathcal{R}}(k)=\frac{k^3}{16\pi^2\epsilon M^2_{\rm Pl}\gamma k_0}e^{-\Delta N}.
\end{eqnarray}
We see that for longer modes the power is scale-dependent
$(\mathcal{P}_{\mathcal{R}}(k)\propto k^3)$ which is not in accordance
with the observations. Hence it has been suggested in
\cite{Martin:2012pe} that these modes which are scale-dependent due to
the effects of CSL mechanism are still outside the horizon and thus
observationally irrelevant. However in this way the observationally
relevant modes will be the ones least affected by CSL and hence it
will not be appropriate to expect localization in field variables for
these modes due to CSL. Thus if one calls for CSL-like mechanism to
explain the classicalization of modes which are observationally
relevant then the scale-dependence of the power spectrum is
inevitable. But, it is interesting to note that in the constant
$\gamma$ model one can consider $k-$dependent $\gamma$ (in which case
all the above derivations hold true) which can cure such discrepancies
in the power spectrum as we will see in the later section.

\section{Effects of scale-dependent `CSL-like' term on inflationary dynamics and squeezing of modes}

In the previous sections we discussed the squeezing of modes in a
generic inflationary scenario as well as in the scenario of Martin
{\it et al.} where `CSL-like' modifications to inflationary dynamics
have been considered with a constant `CSL-like' parameter $\gamma$. In
both these cases we observed that the squeezing of superhorizon modes
happens in the direction of momentum which fails to explain the
macro-objectification of the modes observed in the
field-direction. Also, the scenario of inflation with constant
CSL-like term lacks one essential feature of CSL modifications to
quantum mechanics - namely the `amplification mechanism'.  As the
CSL-like parameter $\gamma$ is introduced as a constant in the
functional Schr\"{o}dinger equation of mode functions (see
Eqn.~(\ref{func-sch-eq})) the rate of localization is the same for all
modes, larger or shorter. On the contrary, in CSL-modified quantum
mechanics larger objects (with larger mass) becomes classical faster
(i.e. their wave-function collapses more rapidly) than shorter objects
which helps keep the microscopic objects in the quantum domain over
astronomical time-scales. Here mass of a system (or a particle while
dealing with single-particle system) has been chosen consciously as
the relevant parameter of efficient collapse. Such a choice is very
natural and driven by our prior knowledge of quantum and classical
systems which can be distinctly discriminated by their mass. In the
same spirit, while applying CSL-like modification to inflationary
dynamics to justify the macro-objectification of the superhorizon
modes, one should expect the modes to behave more classically as they
start crossing the horizon, which indicates that the CSL-like term
$\gamma$ should discriminate between different modes according to
their physical length scales and grow stronger as a mode starts
crossing the horizon during inflation. Hence $\gamma$ should be a
function of length scale (or equivalently of conformal time
$\tau$). Such a choice of scale dependence of $\gamma$ is also a
conscious selection which is based on our prior knowledge of quantum
and classical fields which can be discriminated by their population
density and we know that with increasing length scales modes become
more and more populated. We will see further that similar to the case
in standard inflation, the physical length scale $1/k\tau$ of a mode
$k$ is related to the squeezing parameter $r_k$. Since $r_k$ itself is
directly related to the expectation of occupancy in a particular mode
\cite{Albrecht:1992kf} in the standard inflationary scenario, it seems
natural to expect amplification with respect to the physical size in
the inflationary context as it will correspond to a large occupancy in
that particular mode for which classical behavior is more natural both
from standard QFT and non-relativistic CSL formalism viewpoint.  We
therefore propose a phenomenological ansatz for the form of $\gamma$
as
\begin{eqnarray}
\gamma=\frac{\gamma_0(k)}{(-k\tau)^{\alpha}},
\label{TimeDepG}
\end{eqnarray}
where $\alpha>0$ so that the effects of CSL-like terms dominate as the
modes evolve to cross the horizon and become superhorizon. Secondly,
in the deep subhorizon regime we want the modes to evolve through
standard unitary evolution. Thus, any modification should be
vanishingly small in the extreme subhorizon case in order to obtain
the Bunch-Davies vacuum in that limit.  The above mentioned form of
$\gamma$ seems compatible with this requirement too.

This modification changes the evolution equations for $f_k$ as
\begin{eqnarray}
f_k''+\left(\kappa^2-\frac{a''}{a}\right)f_k=0, 
\label{TimeDepF1}
\end{eqnarray}
where $\kappa^2\equiv k^2-2i\gamma_0(k)(-k\tau)^{-\alpha}$ has now
become time-dependent\footnote{ It is interesting to note that the
  above equation (\ref{TimeDepF1}) resembles with Eqn. (A2) of
  \cite{Martin:2012pe} up to a $k$ dependent $\gamma$ if one takes
  $h(a)$ to be a power law in scale-factor $a$ during inflation.}.
Here too $f_k$ and $\bar f_k$ are two independent solutions of the
above equation. Writing $f_k$ in terms of $u_k$ and $v_k$ as before
(as given in Eqn.~(\ref{f-uv1})) we see that now $u_k$ and $v_k$
satisfy the following evolution equations
\begin{eqnarray}
u_k'&=&-i\kappa u_k+\left(\frac{a'}{a}+\frac{\kappa'}{2\kappa}\right)\bar{v}_k \,,\nonumber\\
v_k'&=&-i\kappa v_k+\left(\frac{a'}{a}+\frac{\kappa'}{2\kappa}\right)\bar{u}_k\,,
\label{uv-time-dep}
\end{eqnarray}
which accordingly change the evolution equations of $r_k,\phi_k$ and
$\theta_k$ as
\begin{eqnarray}
r_k'&=&\left(\frac{a'}{a}+\frac{\kappa'}{2\kappa}\right)\cos{2\phi_k},\nonumber\\
\phi_k'&=&-\kappa-\left(\frac{a'}{a}+\frac{\kappa'}{2\kappa}\right)\coth{2 r_k}\sin{2 \phi_k},\nonumber\\
\theta_k'&=&\kappa+\left(\frac{a'}{a}+\frac{\kappa'}{2\kappa}\right)\sin 2\phi_k\tanh r_k
\label{rtp-eom2}
\end{eqnarray}
where $u_k$ and $v_k$ is parameterized in terms of $r_k,\phi_k$ and
$\theta_k$ as before (see Eqn.~(\ref{uv-rtp})). Rest will remain same
as in the case for constant $\gamma$. One can also check using 
\begin{eqnarray}
\frac{\bar{f}_k'}{\bar{f}_k}=\frac{\bar{u}_k'+v_k'}{\bar{u}_k+v_k}-\frac{1}{2}\frac{\kappa'}{\kappa}
\end{eqnarray}
for time-dependent $\kappa$ and Eqn.~(\ref{uv-time-dep}) that the expression
for $\rm{Re}\,\Omega_k$ also remains the same as given in the
time-independent case. It is difficult to exactly solve these
evolution equations and therefore we will try to solve for the
squeezing parameters approximately as we did before in the case of
constant $\gamma$.
\subsection{Approximate solutions for squeezing parameters}

To solve the evolution equations of the squeezing parameters
approximately we first assume that $r_k^{\rm R}\rightarrow -\infty$ as
$(-k\tau)\rightarrow0$ as we did before which yields
$\coth(2r_k)\rightarrow -1$ in this superhorizon limit. Also we note
that in this limit
\begin{eqnarray}
\left.\frac{\kappa'}{2\kappa}\right|_{-k\tau\rightarrow0}\sim\frac{\alpha}{4}\frac{k}{(-k\tau)}.  
\end{eqnarray} 
This also shows that $\alpha\rightarrow0$ would retrieve the equations
for constant $\gamma$ case. Thus the evolution equations for $r_k$ and
$\phi_k$ simplify in the superhorizon limit as
\begin{eqnarray}
r_k'&=&-\frac{1}{\tau}\left(1+\frac{\alpha}{4}\right)\cos{2\phi_k} ,
\label{RKEqT}\nonumber\\ 
\phi_k'&=&-\kappa-\frac{1}{\tau}\left(1+\frac{\alpha}{4}\right)\sin{2 \phi_k}. 
\label{PhiEqT} 
\end{eqnarray}
Defining $y = -\kappa \tau$ as before the above two equations can be
rewritten as
\begin{eqnarray}
\frac{\partial r_k}{\partial y}&=&-\frac{B}{A}\frac{1}{y}\cos{2 \phi_k},\nonumber\\
\frac{\partial \phi_k}{\partial y}&=&\frac1A-\frac{B}{A}\frac{1}{y}\sin{2 \phi_k},
\end{eqnarray}
where $A\equiv\left(1-\alpha/2\right)$ and
$B\equiv\left(1+\alpha/4\right)$. Making the transformation
$\tan\phi_k=\mathcal{F}_k(y)$ as before the equation for $\phi_k$ can
be written as
\begin{eqnarray}
A \frac{\partial {\mathcal F}_k}{\partial y}-{\mathcal F}_k^2 +\frac{2B}{y} {\mathcal F}_k=1, 
\label{FT1}
\end{eqnarray}
whose solution can be given in terms of Bessel functions of regular
and modified kind as 
\begin{eqnarray}
{\mathcal F}_k(y)=\frac{J_{\frac{1}{2}+\frac{B}{A}}\left(\frac{y}{A}\right)\sec{\frac{(A+B)\pi}{A}}+
  J_{-\frac{1}{2}-\frac{B}{A}}\left(\frac{y}{A}\right)(C+\tan{\frac{B\pi}{A}})}{Y_{\frac{1}{2}-\frac{B}{A}}\left(\frac{y}{A}\right)
  -C
  J_{\frac{1}{2}-\frac{B}{A}}\left(\frac{y}{A}\right)}. 
\label{SolFT1}
\end{eqnarray}
We notice at this point that in the super-horizon limit
($-k\tau\rightarrow 0$) with $2\gamma_0(k)/k^2\gg1$ one has
\begin{eqnarray} 
 y\sim(-k\tau)^{1-\frac{\alpha}{2}}\left(\frac{4 \gamma_0^2(k)}{k^4} \right)^{1/4}\exp{\left[-\frac{i}{2}\tan^{-1}
 \left(\frac{2\gamma_0}{k^2(-k\tau)^{\alpha}}\right)\right]},
 \end{eqnarray}
where $y\rightarrow0$ as $-k\tau\rightarrow0$ if $\alpha<2$.  Thus
considering $0<\alpha<2$ the solution for $\mathcal{F}_k$ simplifies
in the superhorizon limit as
\begin{eqnarray}
\left.{\mathcal F}_k(y)\right|_{y\rightarrow0}\sim\frac{1+\alpha}{y},
\end{eqnarray}
yielding
\begin{eqnarray}
\tan{\phi_k}\approx\frac{1+\alpha}{y}.
\end{eqnarray}
Putting this back in the evolution equation of $r_k$ gives
\begin{eqnarray}
\frac{\partial r_k}{\partial y}\approx\frac{B}{A}\frac{1}{y},
\end{eqnarray}
which yields a solution for $r_k$ in the superhorizon limit as 
\begin{eqnarray}
 r_k\approx\frac BA\ln{y}
\end{eqnarray}
Now, one can explicitly write the real and imaginary parts of $\kappa$, $r_k$
and $\phi_k$ in the superhorizon limit as
\begin{eqnarray}
\kappa^{\rm R} &=&\sqrt{\gamma_0}(-k\tau)^{-\frac\alpha2}\,,\nonumber\\
\kappa^{\rm I} &=& -\sqrt{\gamma_0}(-k\tau)^{-\frac\alpha2}\,,\nonumber\\
r_k^{\rm R}&=&  \ln\left((-k\tau)^{1+\frac\alpha4}\left(\frac{4 \gamma_0^2}{k^4} \right)^{\frac{4+\alpha}{8(2-\alpha)}}\right),\nonumber\\
r_k^{\rm I} &=&-\frac{4+\alpha}{(2-\alpha)}\frac{\pi}{8},\nonumber\\
\tan(\phi_k^{\rm R})&=&\frac{\sqrt{2}}{(-k\tau)^{1-\frac\alpha2}}\left(\frac{2\gamma_0}{k^2}\right)^{-\frac12},\nonumber\\
\tanh\left(\phi_k^{\rm I}\right)&=&\frac{(-k\tau)^{1-\frac\alpha2}}{\sqrt{2}}\left(\frac{2\gamma_0}{k^2}\right)^{\frac12},
\label{real-im-parts-1}
\end{eqnarray}
which shows that with $0<\alpha<2$ one has $r_k^{\rm
  R}\rightarrow-\infty$ as $-k\tau\rightarrow0$ in the superhorizon
limit. This again justifies the assumption we made at the beginning.
\subsection{Wigner function and the macro-objectification of the inflationary modes}

With the solutions of the squeezing parameters in the superhorizon
limit one now can determine the nature of the Wigner function and the
direction of squeezing of the modes. As we discussed earlier, the
squeezing of modes is determined by the variance of exponentials with
field and momentum as its coefficients which is directly related to
${\rm Re}\,\Omega_k$. In the present case, where the CSL-like
parameter $\gamma$ is time-dependent, the explicit form of ${\rm
  Re}\,\Omega_k$ can be given by the one in the case of constant
$\gamma$ as given in Eqn.~(\ref{re-omega-k}). Thus using the
approximate superhorizon solutions of $\kappa$, $r_k$ and $\phi_k$
from Eqn.~(\ref{real-im-parts-1}) one sees that in the limits
$\alpha<2, |k\tau|\ll1$ and $2\gamma_0/k^2\gg1$ the numerator of ${\rm
  Re}\,\Omega_k$ becomes
\begin{eqnarray}
\left[2\sqrt{\gamma_0(k)}(-k\tau)^{-\alpha/2}+\frac{4\sqrt{2}}{k}\gamma_0(k)(-k\tau)^{1-\alpha} \right]\cos{\left(\frac{4+\alpha}{4-2\alpha}\frac{\pi}{2}\right)}
+\frac{2\left(\frac{4\gamma_0^2(k)}{k^4} \right)^{1/2}}{\left(\frac{4\gamma_0^2(k)}{k^4} \right)^{\frac{1}{2}\frac{4+\alpha}{4-2\alpha}}}\frac{k}{(-k\tau)^{1+3\alpha/2}} \nonumber\\
- 2\sqrt{\gamma_0(k)}\sin{\left(\frac{4+\alpha}{4-2\alpha}\frac{\pi}{2}\right)}(-k\tau)^{-\alpha/2}-4\sqrt{2\gamma_0(k)}\left(\frac{4\gamma_0^2(k)}{k^4} \right)^{\frac{1}{4}}
\sin{\left(\frac{4+\alpha}{4-2\alpha}\frac{\pi}{2}\right)}(-k\tau)^{1-\alpha}\nonumber\\
+ 4\sqrt{\gamma_0(k)}\frac{2\left(\frac{4\gamma_0^2(k)}{k^4} \right)^{1/2}}{\left(\frac{4\gamma_0^2(k)}{k^4}
\right)^{\frac{1}{2}\frac{4+\alpha}{4-2\alpha}}}\frac{1}{(-k\tau)^{2\alpha}},\nonumber\\
\end{eqnarray}
whose leading order dependence in $|k\tau|$ will be $(-k
\tau)^{-1-3\alpha/2}$ in the range $0<\alpha<2$. Similarly in the
above limit the denominator of ${\rm Re}\,\Omega_k$ would be
\begin{eqnarray}
4 \left(\frac{4 \gamma_0^2(k)}{k^4} \right)^{1/4}(-k \tau)^{1-\frac{\alpha}{2}}\frac{1}{\sqrt{2}}\cos{\left(\frac{4+\alpha}{4-2\alpha}\frac{\pi}{2}\right)}
+\frac{\left(2+4 \left(\frac{4 \gamma_0^2(k)}{k^4} \right)^{1/4}(-k \tau)^{1-\frac{\alpha}{2}}\frac{1}{\sqrt{2}}\right)}{\left(\frac{4\gamma_0^2(k)}{k^4}
\right)^{\frac{1}{2}\frac{4+\alpha}{4-2\alpha}}}
\frac{1}{2(-k \tau)^{2+\frac{\alpha}{2}}} \nonumber\\
+2\left(1+2\left(\frac{4 \gamma_0^2(k)}{k^4}\right)^{1/4}(-k \tau)^{1-\frac{\alpha}{2}}\frac{1}{\sqrt{2}}\right)\left(\frac{4\gamma_0^2(k)}{k^4}
\right)^{\frac{1}{2}\frac{4+\alpha}{4-2\alpha}}\frac{1}{2(-k \tau)^{2+\frac{\alpha}{2}}},\nonumber\\
\end{eqnarray}
whose leading order dependence in $|k\tau|$ will be $(-k
\tau)^{-(2+\frac{\alpha}{2})}$. Therefore, the leading order behavior
of ${\rm Re}\,\Omega_k$ on super-horizon scales would be
\begin{eqnarray}
{\rm Re}\,\Omega_k\approx \frac k2(-k\tau)^{1-\alpha}\left(\frac{2\gamma_0(k)}{k^2}\right).
\label{scale-dep}
\end{eqnarray}
One can now see that if $0<\alpha<1$ then ${\rm
  Re}\,\Omega_k\rightarrow0$ as $-k\tau\rightarrow0$ on superhorizon
scales which yields a squeezing in the direction of momentum as the
variance in this direction becomes very small which can be seen from
Eqn.~(\ref{wigner-generic}). This also happens in a generic
inflationary scenario as well as in the case where CSL-like correction
is done with a constant $\gamma$ and we see as before that such a case
fails to explain the macro-objectification of the modes observed in
various experiments.

But, the scenario becomes very interesting for the range of $\alpha$
where $1<\alpha<2$. In this particular range one notices that ${\rm
  Re}\,\Omega_k\rightarrow\infty$ as $-k\tau\rightarrow0$ on
superhorizon scales which yields a squeezing in the direction of MS
field variable $\zeta_k$ as the variance in this direction would then
become very small which can be seen from
Eqn.~(\ref{wigner-generic}). 
\begin{figure*}[ht]
\begin{center}
\includegraphics[width=6.5cm, height=5cm]{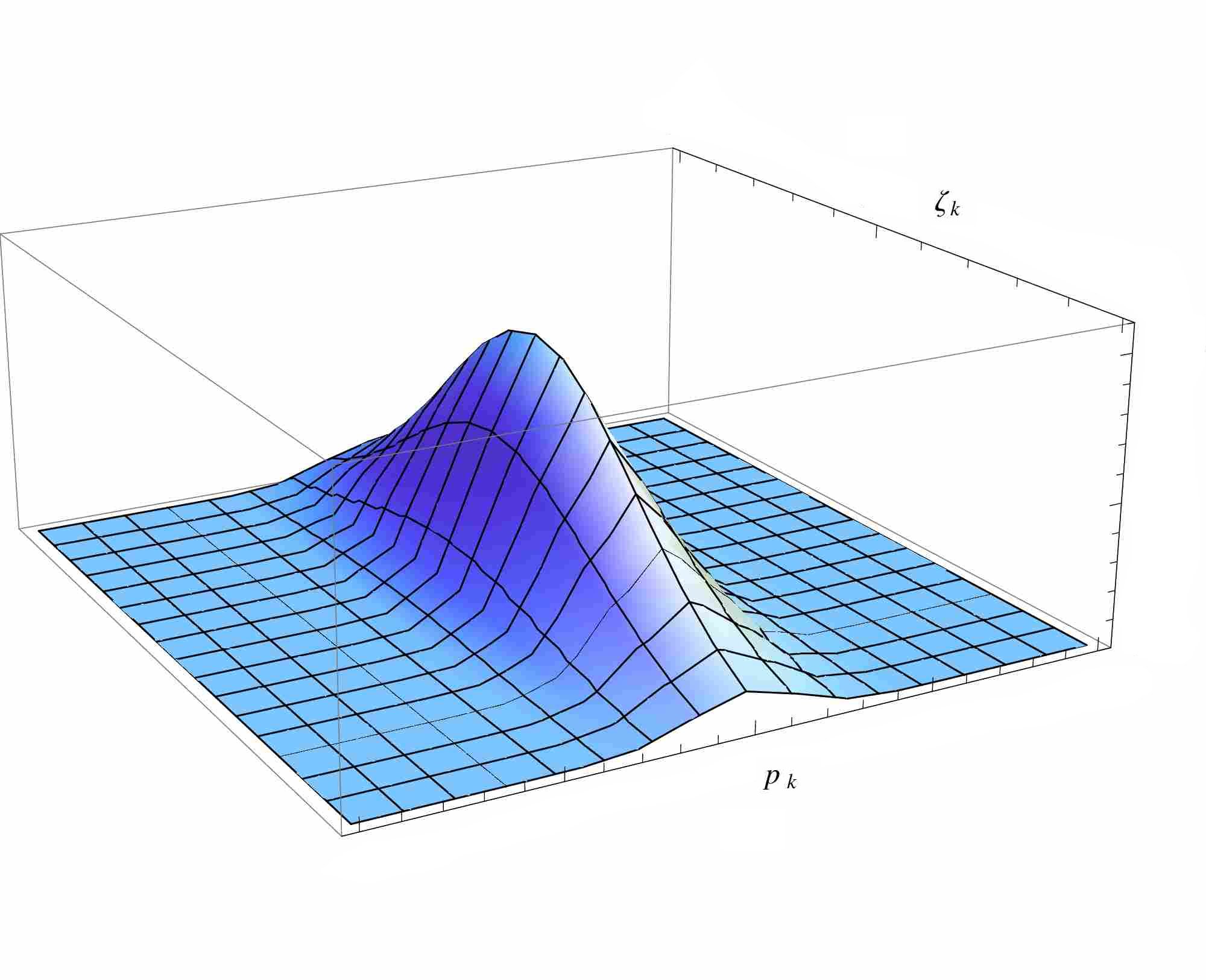}
\includegraphics[width=6.5cm, height=5cm]{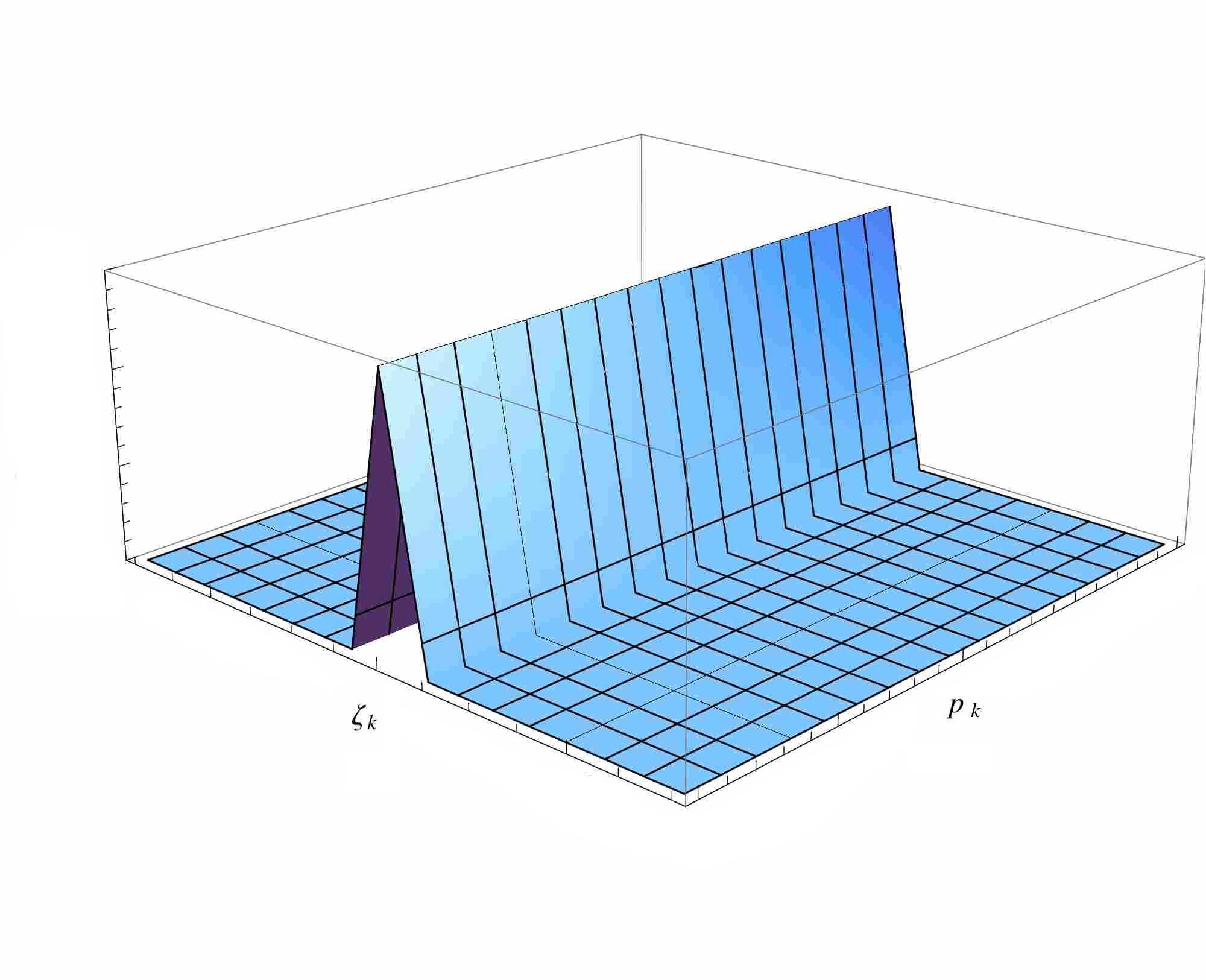}
\end{center}
\caption[]{Plot of Wigner function for $\gamma=100$, $|r_k|=6.5$ and
  $\alpha=0.5$ for the plot in the left panel and $\gamma=100$,
  $|r_k|=6.2$ and $\alpha=1.5$ for the right panel figure. It shows high
  squeezing of a particular mode in the direction of field variable
  for the later case}
\label{wigner3}
\end{figure*}

In Fig.~\ref{wigner3} we see that as long as we have $0<\alpha<1$, the
squeezing remains in the direction of momentum whereas for the range
$1<\alpha<2$ the direction of squeezing changes in favor of the field
variable. This scenario is thus in accordance with the
macro-objectification of the modes as suggested by cosmological
observations without invoking requirement of decoherence and hence
avoiding many-worlds scenario for explaining the observations.

\subsection{Obtaining the scale-invariant Power Spectrum}

We previously saw in the case of constant $\gamma$ that introduction
of CSL-like terms in the inflationary dynamics yield a scale-dependent
power spectrum for the comoving curvature perturbations for the
largest modes which contradicts the precise measurement of CMBR power
spectrum of high-precision observations made by experiments like
WMAP \cite{Hinshaw:2012fq} and PLANCK \cite{Ade:2013uln}. It is thus suggested in \cite{Martin:2012pe} that
these scale-dependent modes affected by constant CSL-like terms are
still superhorizon at present day and thus are not related to these
observations. The shorter modes on the other hand are not affected much
by the CSL-like term $\gamma$ and thus can produce a scale-invariant
power spectrum which is in accordance with the observations. This
suggests that the constant $\gamma$ can spoil the observed scale
invariance of the power spectrum and also suggests that CSL-like
modifications to quantum theory are contradicted by inflationary dynamics. 
However, since now in our case $\gamma$
can depend upon $\tau$ as well as $k$ we can thus try to construct a
scale-invariant power spectrum.

We see from Eqn.~(\ref{power-R}) that the scale-dependence of the
power spectrum comes from the factor $k^3/(a^2{\rm Re}\,\Omega_k)$. It
is also to be noted that this factor depends upon $\tau$ yielding a
time-dependent power spectrum as we also got while considering
constant $\gamma$. In order to remove this $\tau$ dependence in the case of 
constant $\gamma$ the power spectrum was
calculated at the end of inflation using Eqn.~(\ref{end-inf}). We follow the same path
here and calculate the power at the end of inflation.

We will be interested in the parameter regime where $1<\alpha<2$ as in
this regime the macro-objectification of modes can be explained. In
this regime ${\rm Re}\,\Omega_k$ can be approximately written in the
superhorizon limit as
\begin{eqnarray}
{\rm Re}\,\Omega_k \approx \frac k2\left(\frac{2\gamma_0(k)}{k^2}\right)(-k \tau)^{1-\alpha}.  
\label{ReOm}
\end{eqnarray}
Thus in a de Sitter space the scale-dependence of the power spectrum
would be
\begin{eqnarray}
\frac{k^3}{a^2{\rm Re}\,\Omega_k}=\left(\frac{\gamma_0(k)}{k^2}\right)^{-1}(-k\tau)^{1+\alpha}=\left(\frac{\gamma_0(k)}{k^2}\right)^{-1}\left(\frac{k}{k_0}\right)^{1+\alpha}e^{-(1+\alpha)\Delta N},
\end{eqnarray}
where we have used Eqn.~(\ref{end-inf}) for the last equality. In its
present form we can see that if $\gamma_0$ has no dependence upon $k$
then the power would still be scale-dependent. But this can be fixed
by fixing the scale dependence of $\gamma_0(k)$. In doing so we
consider the form of $\gamma_0(k)$ as
\begin{eqnarray}
\gamma_0(k)=\tilde{\gamma_0}\left(\frac{k}{k_0}\right)^{\beta},
\label{gamma_a1}
\end{eqnarray}
where $\tilde{\gamma_0}$ is a constant with no dependence upon $k$ or
$\tau$.
With this we see that the scale-dependence of the power
spectrum of the comoving curvature power spectrum would be
\begin{eqnarray}
\mathcal{P}_{\mathcal{R}}(k)\propto k^{3+\alpha-\beta}
\end{eqnarray}
and thus by setting $\beta=3+\alpha$ one can get a scale-invariant
power spectrum. As we are interested in the range $1<\alpha<2$, this
sets a range for $\beta$ too as $4<\beta<5$. However, as correctly 
pointed out in \cite{Martin:2012pe} the power spectrum remains time-dependent in
such analysis, therefore, making any comparison of the power spectrum at the end of inflation 
with that at recombination non-trivial.

\section{Investigation of Phase Coherence of super-horizon fluctuations under CSL modifications}
\label{ph-coh}

So far, while making modification to inflationary dynamics by adding
CSL-like terms with either constant $\gamma$ or $\gamma$ which has $k$
and $\tau$ dependence, we were concerned about one observational
consequence of inflationary dynamics: scale-invariance of the
anisotropy spectrum of the precisely measured CMBR temperature
fluctuations. On the other hand, it is also very important to note
here that a generic inflationary scenario not only predicts a
scale-invariant power spectrum, but also explains the existence of
sharp peaks and troughs of the CMBR power spectrum which is caused by
the coherent initial phases of all the Fourier modes of curvature
perturbations at horizon re-entry corresponding to a given wave
number. We previously showed (Eqn.~(\ref{phase-coherence}) and
discussion thereafter) how in a generic inflationary scenario the
phase $\delta_k$ of the mode function $f_k$ freezes on super-horizon
scales.  In standard inflationary scenario we will see that it leads
to freezing of the amplitude of the curvature perturbation on
super-horizon scales. Upon re-entry the curvature perturbation begins
to oscillate and it can be shown that all modes corresponding to a
given wave number begin their oscillations with same initial phase
(not to be confused with $\delta_k$) in that case, leading 
to a coherent interference to produce peaks
and troughs in the CMBR power spectrum and a snapshot of it at the last
scattering surface is what we observe today in different experiments
such as WMAP \cite{Hinshaw:2012fq} and PLANCK \cite{Ade:2013uln}. But if
all the Fourier modes of a given length scale had random phases, they would 
have interfered destructively to wash
out all those sharp peaks and troughs of the CMBR spectrum to leave us
simply with a flat spectrum \cite{Dodelson:2003ip, Albrecht:1995bg}.

We have seen before that modifying inflationary dynamics by CSL-like
terms with constant $\gamma$ can yield a scale-dependent power
spectrum for large scales which then contradicts with the
observations. One can preferably keep these modes outside the horizon;
thus making them observationally less important, but the smaller modes
which explain a scale-invariant power spectrum are not much affected
by CSL-like terms and thus their classicality cannot be explained by
modification of inflationary dynamics with CSL mechanism. On the other
hand we have seen that by making the CSL-like parameter $\gamma$
dependent upon $k$ and $\tau$ there exists a parameter range
($1<\alpha<2$ and $4<\beta<5$) where the macro-objectification of
modes as well as a scale-invariant power spectrum can both be
explained simultaneously. In this section we
consider the phase coherence of the superhorizon modes and
investigate whether inflationary dynamics modified by CSL-like
mechanism is also in accordance with phase coherence or whether such
CSL-like modifications of inflationary dynamics can spoil these
patterns we observe in the CMBR spectrum and thus contradict the
observations.

As argued above, phase coherence is related to the freezing of the
amplitude of the curvature perturbation ($|{\mathcal R}_k|$)  with respect to $\tau$.  In standard
inflation this is a trivial exercise to check by the means of the
equation of motion for the variable $f_k$ by substituting $f_k$ as
$R_k\exp{(i\delta_k)}$, where $\delta_k$ carries the overall phase
dependence, since $|{\mathcal R}_k|$ is given by
the quantity $|f_k/a |$.  The equations for its real and imaginary parts can be
separated as \cite{Goswami:2010qu} 
\begin{eqnarray}
\frac{R_k^{''}}{R_k}+\left[k^2-\frac{a^{''}}{a}-(\delta_k')^2\right]&=&0, \label{Real}\\ 
\delta_k^{''}+2\left(\frac{R_k'}{R_k}\right)\delta_k'&=&0,\label{Imaginary}
\end{eqnarray} 
respectively. From Eqn.~(\ref{Imaginary}) we can see that
$\delta_k'=0$ is a fixed point of the equation and $\delta_k'$
approaches this value asymptotically, i.e. for $|k \tau|\rightarrow
0$. Therefore, $R_k\rightarrow a$ and hence $ |f_k/a|$ is constant
resulting in the phenomenon of phase coherence.

But we note here that similar analysis of phase coherence in
the case of CSL-modified inflationary dynamics would not hold as in
such a case it is difficult to identify the parameter $f_k$ as the mode
function as has been discussed before. But the requirement of phase
coherence is related to freezing of amplitude of ${\mathcal R}_k$ on
super horizon scales. We see from Eq.~(\ref{power-R}) that in
Schr\"{o}dinger picture analysis the amplitude of curvature
perturbations varies as 
\begin{eqnarray}
|{\mathcal R}_k|\propto \frac{1}{\left(a^2{\rm Re}\,\Omega_k\right)^{\frac12}}.
\end{eqnarray}
As the behavior of $a^2{\rm Re}\,\Omega_k$ is known for superhorizon
modes for all the cases we discussed above, one can determine whether
the amplitude of comoving curvature perturbation freezes on
superhorizon scales in each such case or not. Let us analyze the
phenomena of phase coherence case by case :

\noindent 1. \underline{Constant $\gamma$ modification for larger modes }:

\noindent We see from Eq.~(\ref{large}) that in such a case 
\begin{eqnarray}
\left(a^2{\rm Re\,}\Omega_k\right)^{\frac12}\propto 1/\sqrt{-\tau},
\end{eqnarray}
which shows that 
\begin{eqnarray}
\frac{d|{\mathcal R}_k|}{d\tau}\propto 1/\sqrt{-\tau},
\end{eqnarray}
indicating that on superhorizon scales ($-k\tau\rightarrow0$) the
amplitude grows and would not freeze. Thus such modes can not lead to
phase coherence of the CMBR spectrum. We recall that these modes also
violate the scale invariance of the power spectrum and thus are 
inconsistent with observation.

\noindent 2. \underline{Constant $\gamma$ modification for smaller modes }:

\noindent In this case we see from Eq.~(\ref{small}) that $a^2{\rm Re}\,\Omega_k$ is
a constant in $\tau$ which indicates that $|{\mathcal R}_k|$ does freeze on
superhorizon scales and thus can give rise to the observed phase
coherence of the CMBR spectrum. We also recall that such modes can
yield a scale invariant spectrum and thus are observationally
consistent. But as these modes are the ones least affected by the collapse
operator and lead to a Wigner function squeezed in the momentum
direction, rather than field direction, they do not satisfy the macro-objectivity criterion.

\noindent 3. \underline{Modification with scale dependent $\gamma$ case }:

\noindent We see from Eq.~(\ref{scale-dep}) that 
\begin{eqnarray}
\left(a^2{\rm Re}\,\Omega_k\right)^{\frac12}\propto \frac{1}{(-\tau)^{\frac{1+\alpha}{2}}},
\end{eqnarray}
which yields the evolution of amplitude of curvature perturbations on
superhorizon scales as
\begin{eqnarray}
\frac{d|{\mathcal R}_k|}{d\tau}\propto (-\tau)^{(\alpha-1)/2}.
\end{eqnarray}
This indicates that for $\alpha>1$ the amplitude freezes on
superhorizon scales and thus such modes can give rise to the observed
phenomena of phase coherence in the CMBR spectrum. We also recall that
this bound on $\alpha$ is also in accordance with
macro-objectification of modes. Thus once $1<\alpha<2$ one can obtain macro-objectification
of the modes and can also remain consistent with observations.

\section{Conclusions}
In this work we have analyzed an alternative scheme of dealing with
the issue of classicalization of inflationary perturbations. This
mechanism, although phenomenological at best, derives its motivation
from a similar proposed modification to quantum theory, known as
Continuous Spontaneous Localization, which addresses the issue of
classicalization of macroscopic systems in non-relativistic
physics. Although this scheme is not fully developed and still lacks a
relativistic generalization, one can make an attempt to implement a
similar stochastic correction in the Schr\"{o}dinger representation of
the field theoretic description. Martin {\it et al.}
\cite{Martin:2012pe} recently took a step in this direction by
considering a CSL analog with constant $\gamma$ parameter. But it
fails to capture the essential feature of such schemes, namely the
amplification mechanism.  Moreover, such a constant $\gamma$ model
also leads to a distortion of the scale-invariance of power spectrum
and so is in conflict with cosmological observations.

We take this approach further and consider a variable $\gamma$
model. We expect such a generalization to make the wavefunction evolve
into one of the eigenfunctions of field variable. This should reflect
itself in the squeezing of Wigner function along the field
direction. The standard inflationary scenario squeezes the Wigner
function along the momentum direction. Also, neither a small constant $\gamma$ nor any modification of
the form of $\gamma(k)$ can yield squeezing in the field direction and
hence fail to explain `single-outcome' of the measured modes. Although one
can take arbitrarily large $\gamma$ to cure for the squeezing direction, but 
our result shows that such models will still be at variance with phase coherence observed in the CMBR profile. However
one can attempt a generalization where the CSL strength parameter is
made time dependent so that the strength of correction will depend on
the physical length scales of concerned modes. We see that although
the most general class of `collapsing models' will distort the power
spectrum, there does exist a subfamily of these models which can generate
scale invariance in accordance with the data without constraining the
model further.  It is to be noted here that in
order to achieve scale-invariance of the spectrum we have exploited the
arbitrariness of $k$ dependence of $\gamma$. It has so far no
justification from the point of view of field theory and rather it is
purely phenomenological. But until a relativistic generalization of
CSL is achieved, this arbitrariness, we believe, is difficult to avoid.

In addition to scale-invariance of the power spectrum, the CMBR data
also suggests the phase coherence of initial density perturbations
which manifests itself in the acoustic peaks of the CMBR map. This
observation also constrains one of the parameters, namely $\alpha$ 
to be greater than 1, which tallies with the requirement of macro-objectification.
 Thus such a model can account for
cosmological observations such as scale-invariance of the primordial
spectrum and the existence of acoustic peaks in the CMBR while
providing a mechanism for macro-objectification. This also means that
such bounds on the model parameter $\alpha$ should be respected by any
suitable relativistic modification of CSL to be compatible with
cosmology.

One important aspect of such CSL-like analysis in the Fourier basis is
the non-conservation of energy. CSL formalism generically suffers with
an ever-present non-conservation of energy \cite{Bassi:2003gd} in the
infinite temperature thermal bath model. As discussed in
\cite{Martin:2012pe}, a mode-by-mode analysis of inflationary
fluctuations will make the expectation of the Hamiltonian diverge. For
constant $\gamma$ model the divergence is $k^3$ while for our model
the divergence is quite severe with a $k^n$ dependence where
$n>6$. Such a strong increase in the energy density, even after
regularization, can possibly lead to back-reaction and will be worth
studying in future.

\section*{Acknowledgements}

This work was supported by a grant from the John Templeton Foundation
(ID 20768). The support of the Foundational Questions Institute is gratefully acknowledged.
 We are grateful to Angelo Bassi for suggesting this
problem, namely the possible application of the CSL mechanism to the
issue of classicality of inflationary density perturbations.  We would
like to thank R. Khatri, C. Kiefer, J. Martin, P. Peter and V. Vennin
for insightful correspondence. TPS acknowledges useful discussions
with Angelo Bassi, Marie-Noelle Celerier and Daniel Sudarsky. 


\end{document}